\documentclass[onecolumn,sort&compress,numbers]{els-mrw} 

\usepackage{amsmath,amssymb,amsfonts,amsthm,graphicx}
\usepackage{txfonts}
\usepackage{helvet}

\usepackage[utf8]{inputenc}
\usepackage[english]{babel}
\usepackage[dvipsnames]{xcolor}
\usepackage[normalem]{ulem}
\usepackage[
colorlinks=true,
linkcolor=blue,
breaklinks=true,
urlcolor=magenta,
citecolor=blue]{hyperref}

\usepackage{bm,color,xcolor,amsfonts,amsmath,amssymb,epsfig,esint,orcidlink}

\usepackage{mathtools,slashed,braket}

\newcommand{\vp}{\varphi}

\newcommand{\be}{\begin{equation}}
\newcommand{\ee}{\end{equation}}
\newcommand{\bea}{\begin{eqnarray}}
\newcommand{\eea}{\end{eqnarray}}
\newcommand{\beas}{\begin{eqnarray*}}
\newcommand{\eeas}{\end{eqnarray*}}
\newcommand{\ds}{\displaystyle}
\newcommand{\vep}{{\bm p}}

\newcommand{\veq}{{\bm q}}

\newcommand{\ven}{{\bm n}}
\newcommand{\vex}{{\bm x}}
\newcommand{\vey}{{\bm y}}

\newcommand{\vesig}{{\bm \sigma}}

\newcommand{\LL}{\mathcal{L}}

\newcommand{\Ps}{P}

\def\vec#1{\boldsymbol{#1}}

\newcommand{\gt}{\chi}

\newcommand{\power}{\alpha}

\graphicspath{{Figs.dir/}}

\begin{document}



\chapter{Chiral symmetry and its breaking}\label{chap1}

\author[1]{Alexey Nefediev}%

\address[1]{\orgname{Universit\"at Bonn}, \orgdiv{Helmholtz-Institut f\"ur Strahlen- und Kernphysik}, \orgaddress{D-53115 Bonn, Germany}}


\maketitle

\begin{abstract}[Abstract]
In addition to fundamental symmetries playing a crucial role for establishing the Standard Model of fundamental interactions, approximate symmetries provide essential insight into the respective phenomena and shed light on the underlying physics. Here we give a brief pedagogical introduction to chiral symmetry as an approximate but still rather accurate symmetry of strong interactions and its spontaneous breaking in the vacuum of Quantum Chromodynamics. Special attention is paid to a microscopic picture of this phenomenon and understanding a dual nature of the chiral pion that is the Goldstone boson related to spontaneous breaking of chiral symmetry and the lowest pseudoscalar quark-antiquark state in the spectrum of hadrons simultaneously.
\end{abstract}

\begin{keywords}
chiral symmetry\sep strong interactions\sep spontaneous breaking of symmetry \sep quark flavour \sep pions \sep quark models
\end{keywords}

\begin{glossary}[Nomenclature]
	\begin{tabular}{@{}lp{34pc}@{}}
		BCS & Bardeen, Cooper, and Schrieffer\\
		GNJL & Generalised Nambu--Jona-Lasinio\\
		NJL & Nambu--Jona-Lasinio \\
		PCAC & partial conservation of axial-vector current\\
		QCD & Quantum Chromodynamics\\
		QED & Quantum Electrodynamics\\
		QFT & Quantum Field Theory\\
		SBCS & Spontaneous breaking of chiral symmetry\\
		SM & Standard Model
	\end{tabular}
\end{glossary}

\section*{Objectives}
\begin{itemize}
\item Sec.~\ref{sec:int} contains a general introduction.
\item In Sec.~\ref{sec:LR}, general concepts of the right- and left-handed fermions are introduced.
\item In Sec.~\ref{sec:vecax}, vector and axial currents and their conservation laws (Ward identities) are derived and discussed.
\item In Sec.~\ref{sec:Nf}, the case of the number of fermion flavours $N_f\geqslant 2$ is studied and chiral symmetry is introduced.
\item In Sec.~\ref{sec:QCD} chiral symmetry is introduced to strong interactions.
\item In Sec.~\ref{sec:CS}, different realisations of chiral symmetry are discussed and its spontaneous breaking is explained and exemplified.
\item In Sec.~\ref{sec:models}, a microscopic picture of spontaneous breaking of chiral symmetry in the vacuum is presented based on quantum-field-theory-inspired quark models.
\item In Sec.~\ref{sec:pion}, we discuss the physics of the chiral pion that emerges as the Goldstone boson of SBCS but can also be understood as the lightest pseudoscalar quark--antiquark state in the spectrum of hadrons.
\end{itemize}

\section{Introduction}
\label{sec:int}

Symmetries are the cornerstone of the contemporary theory of fundamental interactions known as the Standard Model (SM). For instance, invariance of the action of the theory with respect to the transformations from the Lorentz group and the relevant gauge group is an essential prerequisite that allows one to identify the proper building blocks for the theory in terms of the fields and constrain possible interactions between them. Such symmetries are regarded as fundamental and there exists a vast literature on the subject --- see, for example, the books \cite{Coleman:1985rnk,Georgi:1999wka,Georgi:2000vve,Quigg:2013ufa,Weinberg:1995mt} as well as many others. Chiral symmetry addressed below does not belong to this class and appears to be an approximate symmetry of the action if the masses of the fermions present in the theory are small enough. Nevertheless, this symmetry and the physics behind it play an essential role in our understanding of the fundamental interactions and classification of hadronic states predicted by the theory and observed experimentally.
The aim of this essay is a pedagogical introduction to the basic concepts related to chiral symmetry and its spontaneous breaking in the vacuum of Quantum Chromodynamics (QCD).
Special emphasis is put on a microscopic picture of this phenomenon that can be drawn employing suitable quark models. In the spotlight of the discussion below is an interpretation of the dual nature of the chiral pion (more generally, the octet of the light pseudoscalar mesons in the hadronic spectrum of QCD) that appears as the Goldstone boson of spontaneous breaking of chiral symmetry (SBCS) in the QCD vacuum and at the same time can be understood as the lightest pseudoscalar quark-antiquark state in the spectrum of hadrons built on top of the chirally broken vacuum.

The selection of the material and a particular way of its presentation below
reflect personal preferences of the author. The cited materials on such a well studied and documented phenomenon widely discussed in the literature
should be regarded not as a complete list of sources on the subject but as a possible starting point for a new reader. A deeper insight into the discussed phenomena and related physics can be gained from classical textbooks such as \cite{Itzykson:1980rh,Cheng:1984vwu,Weinberg:1995mt,Peskin:1995ev} and many others. See also the chapter on Chiral Perturbation Theory \cite{Meissner:2024ona}.

\section{Right- and left-handed fermions}
\label{sec:LR}

We start from Quantum Electrodynamics (QED) --- the simplest field theory entering the SM.
The fields of matter in QED such as the electron, muon, and $\tau$-lepton are spin-$\frac12$ fermions which, in a relativistic theory, can be described by two not equivalent two-component spinors tagged below as $\xi$ and $\eta$. While rotations in the three-dimensional space and Lorentz boosts (passing over to a moving reference frame) do not mix these spinors, the $P$-transformation (inversion of the three-dimensional space, $P:\vex\to-\vex$) mixes them.
For this reason, it is convenient to consider both spinors simultaneously and deal with a four-component complex-valued quantity known as ``bispinor'',
\be
\psi=\left(\xi\atop \eta\right).
\label{bispinordef}
\ee
In terms of bispinors, the Lagrangian of a free fermionic field (Dirac Lagrangian) reads\footnote{By adding a full derivative, irrelevant for the equations of motions, this Lagrangian can be brought to a more standard form ${\cal L}_D=\bar{\psi}(i\slashed{\partial}-m_0)\psi$, where $\slashed{\partial}\equiv\gamma^\mu\partial_\mu$.}
\be
{\cal L}_D=\frac{i}{2}\bar{\psi}\gamma^\mu\partial_\mu\psi-\frac{i}{2}(\partial_\mu\bar{\psi})\gamma^\mu\psi-m_0\bar{\psi}
\psi,
\label{DiracLag}
\ee
with $\bar{\psi}=\psi^\dagger\gamma^0$ for the Dirac-conjugated field, $m_0$ for the current fermion mass, and $\gamma^\mu$ for the four anticommuting Dirac matrices $4\times 4$ ,
\be
\{\gamma^\mu,\gamma^\nu\}=2g^{\mu\nu}\hat{I}.
\ee
An explicit form of the Dirac matrices depends on the representation used. In particular, in the Weyl representation, they read
\be
\gamma^0_{\mbox{\tiny $W$}}=\left(\begin{array}{cc}0&\hat{1}\\\hat{1}&0\end{array}\right),\qquad
{\bm\gamma}_{\mbox{\tiny $W$}}=\left(\begin{array}{cc}0&-\vesig\\
\vesig&0\end{array}\right),
\label{gammasWeyl}
\ee
with $\sigma$'s for the $2\times 2$ Pauli matrices.\footnote{The choice of the signs of the Dirac matrices is a convention that may differ from textbook to textbook. For example, in the book \cite{Peskin:1995ev}, the spatial Dirac matrices ${\bm\gamma}_{\mbox{\tiny $W$}}$ are chosen with an opposite sign. Although a different sign convention cannot affect observables, intermediate formulae may look differently.}
In what follows, the subscript $W$ will be omitted for brevity.

The Euler--Lagrange equations of motion for $\psi$ and the Dirac-conjugated field $\bar{\psi}$ derived from the stationary action principle read
\be
\frac{\partial\LL}{\partial\bar{\psi}}-\partial_\mu\frac{\partial\LL}{\partial(\partial_\mu\bar{\psi})}=0,\qquad
\frac{\partial\LL}{\partial\psi}-\partial_\mu\frac{\partial\LL}{\partial(\partial_\mu\psi)}=0.
\label{ELeq}
\ee
Then, using the Lagrangian in Eq.~\eqref{DiracLag}, the free field $\psi$ can be found to obey the famous Dirac equation,
\be
(i\slashed{\partial}-m_0)\psi=0,
\label{Diraceq}
\ee
which, with the help of the explicit form of the Dirac matrices in Eq.~\eqref{gammasWeyl}, is written in components as
\be
i\left(\frac{\partial}{\partial t}+(\vesig\cdot{\bm\nabla})\right)\xi=m_0\eta,\qquad
i\left(\frac{\partial}{\partial t}-(\vesig\cdot{\bm\nabla})\right)\eta=m_0\xi.
\label{DiraceqWeyl}
\ee
The degenerate system of equations \eqref{DiraceqWeyl} possesses a nontrivial stationary solution $\psi\propto e^{-iEt+i\vep\vex}$, with $E$ and $\vep$ the fermion energy and 3-momentum, respectively, only if the determinant of the corresponding matrix vanishes,
\be
\mbox{det}\left(
\begin{matrix}
E-(\vesig\cdot\vep) & -m_0\\
-m_0 & E+(\vesig\cdot\vep)
\end{matrix}
\right)=0\qquad\Longrightarrow\qquad
E=\pm\sqrt{\vep^2+m_0^2},
\label{Ep}
\ee
with the plus and minus sign for the fermion and antifermion, respectively.
Remarkably, in the employed Weyl representation for the Dirac matrices, the upper and lower components of the bispinor $\psi$ couple to each other through the mass terms on the right-hand sides of the equations in Eq.~\eqref{DiraceqWeyl}. Then, for a massless fermion with $m_0=0$ and $E=|\vep|$, the two equations decouple and take a particularly simple form,
\be
\frac12(\vesig\cdot\ven)\xi=\frac12\xi,\qquad
\frac12(\vesig\cdot\ven)\eta=-\frac12\eta,
\label{Diraceqxieta}
\ee
where $\ven=\vep/p$ and $\frac12(\vesig\cdot\ven)$ can be readily recognised as the helicity operator that returns the projection of the spin of the fermion on its momentum. Then it is easy to see from Eq.~\eqref{Diraceqxieta} that the spinors $\xi$ and $\eta$ are no more than the eigenfunctions of the helicity operator that correspond to the eigenvalues $s=\pm\frac12$, respectively. Notice also that
the helicity operator commutes with the Dirac Hamiltonian,
\be
{\cal H}_D={\bm\alpha}\cdot\vep+\beta m_0,\qquad {\bm\alpha}=\gamma_0{\bm\gamma},\qquad \beta=\gamma_0,
\label{DiracHam}
\ee
which follows from the Dirac equation \eqref{Diraceq} written in the form of a Schr{\"o}dinger equation $i\partial\psi/\partial t={\cal H}_D\psi$.
Therefore, helicity is a constant of motion and as such can be employed to specify the spin state of the fermion. Notice that, for a massive fermion, the external observer can choose a reference frame moving in the same direction as the fermion but faster than it. In this reference frame, the momentum of the fermion reverses the sign, so its helicity flips. In the meantime, since a massless fermion travels with the speed of light, there exists no reference frame moving faster than it. Therefore, the helicity of a massless fermion is Lorentz invariant although the helicity of a massive fermion is not.
For the fermion helicity taking the two allowed values, $s=\pm\frac12$, the relations in Eq.~\eqref{Diraceqxieta} imply that
\be
\psi_\frac12=\left(\xi_\frac12\atop 0\right),\qquad
\psi_{-\frac12}=\left(0\atop \eta_{-\frac12}\right),
\label{upanddown}
\ee
respectively, while $\xi_{-\frac12}=\eta_{\frac12}=0$.

\begin{figure}[t]
\centering
\begin{tabular}{ccc}
\includegraphics[width=0.25\textwidth]{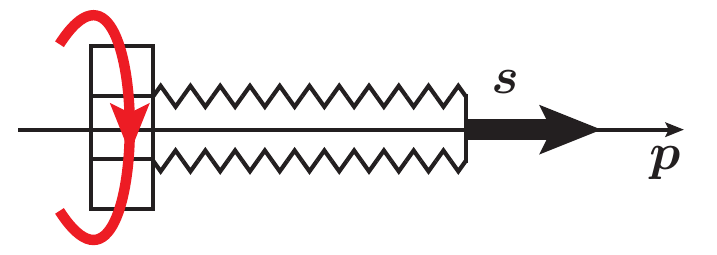}&\hspace*{0.08\textwidth}&
\includegraphics[width=0.25\textwidth]{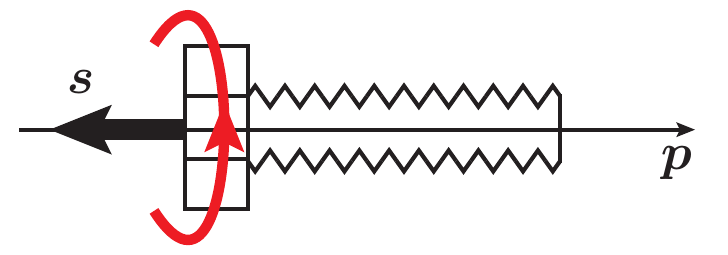}\\
(a)&&(b)
\end{tabular}
\caption{A screw (a) rotated clockwise representing a right-handed object with positive ``helicity'' and (b) rotated counterclockwise representing a left-handed object with negative ``helicity''.}
\label{fig:screw}
\end{figure}

Consider now a simple mechanical system depicted in Fig.~\ref{fig:screw} and consisting of a screw with a right thread rotated by a screwdriver clockwise (to the right) or counter clockwise (to the left). In the former case,
the ``spin'' associated with the rotation is aligned with the momentum thus resulting in a positive ``helicity''. In the latter case, the ``spin'' is counter aligned with the momentum, so the object has negative ``helicity''.
Extending this classical analogy to fermions one can say that a massless fermion with a positive or negative helicity can be regarded as right- or left-handed, respectively. This way we arrive at the notion of chirality (from the Greek word ``$\chi\epsilon\iota\rho$'' meaning an ``arm'') as an intrinsic property of massless fermions. To put this notion on a formal ground notice that
an auxiliary (fifth) Dirac matrix,
\be
\gamma_5=\gamma^5=i\gamma^0\gamma^1\gamma^2\gamma^3=-i\gamma_0\gamma_1\gamma_2\gamma_3=\left(\begin{array}{cc}1&0\\0&-1\end{array}\right),
\label{gamma5}
\ee
commutes with the Dirac Hamiltonian in Eq.~\eqref{DiracHam} for $m_0=0$ and, therefore, so do the projectors,\footnote{Note that different textbooks and original papers may use different sign conventions for the matrix $\gamma_5$, so the signs in the projectors can also alternate.}
\be
P_R=\frac12(1+\gamma_5)=
\begin{pmatrix}
1&0\\0&0
\end{pmatrix},
\qquad
P_L=\frac12(1-\gamma_5)=
\begin{pmatrix}
0&0\\0&1
\end{pmatrix}.
\label{PRL}
\ee
Then the fermion field $\psi$ can be identically decomposed into its right- and left-handed components,
\be
\psi=\psi_R+\psi_L,\qquad
\psi_R=P_R\psi=\left(\xi\atop 0\right),\qquad
\psi_L=P_L\psi=\left(0\atop \eta\right),
\label{psiRL}
\ee
and it is easy to see that the helicity states in Eq.~\eqref{upanddown} are eigenstates of the projectors $P_{R,L}$ above --- a right(left)-handed fermion has a positive(negative) helicity. Thus, a massless fermion possesses both definite helicity and chirality at the same time. This correlation is especially transparent in the Weyl representation for the Dirac matrices which, for this reason, is also known as the chiral representation.
It should also be clear from the consideration above that the chirality of a massless fermion is not only Lorentz invariant by construction but is also a constant of motion. The latter property is, however, spoilt by the mass term in the Lagrangian \eqref{DiracLag}. Indeed, it is easy to verify that
\be
\bar{\psi}\gamma^\mu\psi=\bar{\psi}_R\gamma^\mu\psi_R+\bar{\psi}_L\gamma^\mu\psi_L,
\qquad
\bar{\psi}\psi=\bar{\psi}_R\psi_L+\bar{\psi}_L\psi_R,
\label{vecscalRL}
\ee
so the right- and left-handed components of the fermionic field are decoupled from each other in the kinetic term of the Lagrangian but such decoupling does not hold for the mass term.
The second relation in Eq.~\eqref{vecscalRL} implies that the mass term in the Lagrangian \eqref{DiracLag} mixes different chiralities and, therefore, chirality is not a constant of motion for a massive fermion --- the latter can change its chirality over time.
Interestingly, a seemingly artificial and trivial decomposition of the fermionic field in Eq.~\eqref{psiRL} into its right- and left-handed components has far-reaching consequences for the physics of fundamental interactions. In particular, for the reasons that still remain to be understood, right- and left-handed fermions enter the Lagrangian of weak interactions in a severely different way, which entails \emph{inter alia} violation of the $P$-invariance in weak processes. Further details can be found in the dedicated sources on weak interactions while below we shall concentrate on the implications for strong interactions.

\section{Vector and axial-vector currents}
\label{sec:vecax}

After a brief introduction in the previous section to the properties of a Dirac spin-$\frac12$ particle, we come to a discussion of the symmetries inherent to its action. It is easy to convince oneself that the Lagrangian in Eq.~\eqref{DiracLag} is invariant under the global gauge transformation,
\be
\psi\to e^{-i\gt}\psi,\qquad\bar{\psi}\to \bar{\psi}e^{i\gt},
\label{gt}
\ee
with some constant $\chi$. The full set of transformations in Eq.~\eqref{gt} constitutes a unitary group conventionally tagged as $U(1)$. Since the elements of the group commute with each other, the group is identified as Abelian.
According to the Noether theorem, the global symmetry of the action \eqref{gt} should entail the existence of a conserved current. Indeed, for $\chi\ll1$,
the gauge transformation in Eq.~\eqref{gt} takes a form
\be
\psi\to\psi+\delta\psi,\qquad \bar{\psi}\to\bar{\psi}+\delta\bar{\psi},\quad\mbox{with}\quad
\delta\psi=-i\gt\psi,\quad \delta\bar{\psi}=
i\gt\bar{\psi},
\label{deltapsi}
\ee
so, under this infinitesimal gauge transformation, the Lagrangian in Eq.~\eqref{DiracLag} acquires an additional term,
\be
\delta\LL=\frac{\partial\LL}{\partial\psi}\delta\psi+\delta\bar{\psi}\frac{\partial\LL}{\partial\bar{\psi}}+
\frac{\partial\LL}{\partial(\partial_\mu\psi)}\delta(\partial_\mu\psi)+
\delta(\partial_\mu\bar\psi)\frac{\partial\LL}{\partial(\partial_\mu\bar\psi)}
=i\gt\partial_\mu\left[\bar{\psi}\frac{\partial\LL}{\partial(\partial_\mu\bar\psi)}
-\frac{\partial\LL}{\partial(\partial_\mu\psi)}\psi\right]=
\gt\partial_\mu j^\mu,
\ee
where we used the equations of motion in Eq.~\eqref{ELeq} and defined
\be
j^\mu=\bar\psi\gamma^\mu\psi.
\label{jmu}
\ee
Gauge invariance of the Lagrangian in Eq.~\eqref{DiracLag} implies that
$\delta\LL$ must vanish for any $\gt$ and, therefore,
\be
\partial_\mu j^\mu=0,
\label{veccurrcons}
\ee
which can be readily identified as a fundamental conservation law for the vector electric current (for convenience defined up to the elementary electric charge $e$).\footnote{Here we refrain from dwelling on the issues, irrelevant for this discussion, related to the quantisation and renormalisation of the corresponding theory and, in particular, a proper definition of the electric current through a normal ordering of the fermionic creation and annihilation operators or its symmetrisation with respect to the particles and antiparticles; the corresponding material can be found in textbooks on QED.} For this reason, the corresponding group of the gauge transformations in Eq.~\eqref{gt} is often tagged as $U(1)_V$ to pinpoint its vector nature. Relation \eqref{veccurrcons} can also be derived straightforwardly by applying the derivative $\partial_\mu$ to the current \eqref{jmu} constructed from the fields $\psi$ and $\bar{\psi}$ and then, again, using their equations of motion in Eq.~\eqref{ELeq}. Notice that the decomposition in the first relation in Eq.~\eqref{vecscalRL} implies that the vector current $j^\mu$ naturally splits into individual contributions from the right- and left-handed fermions,
\be
j^\mu=j^\mu_R+j^\mu_L,\qquad j^\mu_R=\bar\psi_R\gamma^\mu \psi_R=\bar\psi\gamma^\mu P_R\psi,\qquad
j^\mu_L=\bar\psi_L\gamma^\mu \psi_L=\bar\psi\gamma^\mu P_L\psi,
\label{jmuLR}
\ee
with the right- and left-hand projectors previously introduced in Eq.~\eqref{PRL}.
Remarkably, the inclusion of the matrix $\gamma_5$ into consideration allows one to additionally build a pseudoscalar and axial-vector bilinear combinations of the fermion fields as
\be
\Ps=i\bar{\psi}\gamma_5\psi,\qquad
j_5^\mu=\bar{\psi}\gamma^\mu\gamma_5\psi,
\label{Psj5}
\ee
respectively. Applying $\partial_\mu$ to $j_5^\mu$ and, as before, using the equations of motions in Eq.~\eqref{ELeq} and the definitions in Eq.~\eqref{Psj5}, one finds
\be
\partial_\mu j_5^\mu=2m_0\Ps,
\label{axveccurrconsclass}
\ee
which turns to the axial current conservation law in the limit of $m_0=0$. In other words, for a massless fermion, not only the vector current $j^\mu$ in Eq.~\eqref{jmu} is conserved but so is the axial current $j^\mu_5$ in Eq.~\eqref{Psj5}, too. It implies that the action of a massless Dirac particle is invariant
under both the gauge transformation in Eq.~\eqref{gt} and the transformation of the form
\be
\psi\to e^{-i\gt'\gamma_5}\psi,\qquad\bar{\psi}\to \bar{\psi}e^{-i\gt'\gamma_5},
\label{gt5}
\ee
with an arbitrary constant $\gt'$. Then, combining the fields $\psi$ and $\gamma_5\psi$ into $\psi_L$ and $\psi_R$ as introduced in Eq.~\eqref{psiRL} above, we conclude that, for a massless fermion, its right- and left-handed components can be transformed independently from each other without changing the action. In other words, the theory with massless fermions possesses a symmetry $U(1)_L\otimes U(1)_R$ (see also the discussion of the charges in the next section). It should be stressed, however, that the consideration above was essentially classical and did not take into account quantum fluctuations. Meanwhile, in the presence of the electromagnetic field, while the vector current conservation law (Ward identity) in Eq.~\eqref{veccurrcons} remains intact thus marking the electric charge conservation as a fundamental law of nature, the classical axial-vector Ward identity in Eq.~\eqref{axveccurrconsclass} is modified,
\be
\partial_\mu j_5^\mu=2m_0\Ps+\frac{e^2}{8\pi^2}F_{\mu\nu}\tilde{F}^{\mu\nu},
\label{axveccurrcons}
\ee
where $\tilde{F}_{\mu\nu}=\frac12\epsilon^{\mu\nu\lambda\rho}F_{\lambda\rho}$ is the dual field tensor. The last term on the right-hand side is known as the axial (or ABJ for the names of Adler, Bell, and Jackiw who observed it in 1969) anomaly \cite{Adler:1969gk,Bell:1969ts}. Equation \eqref{axveccurrcons} implies that, at the quantum level, the axial-vector current is never conserved even for massless fermions since the corresponding symmetry is explicitly broken by the anomaly.\footnote{A well-known consequence of Eq.~\eqref{axveccurrcons} is the two-photon decay of the neutral pion --- see, for example, the related discussion in the chapter on pion decays \cite{Bryman:2025pet}.}
In other words, the classical symmetry of the action $U(1)_L\otimes U(1)_R$ is explicitly broken by the quantum anomaly down to the vectorial $U(1)_V$,
\be
U(1)_L\otimes U(1)_R\mathop{\longrightarrow}^{\rm Anomaly} U(1)_V.
\ee

\section{Chiral symmetry for $N_f$ fermion flavours}
\label{sec:Nf}

The symmetry scheme outlined in the previous section is relevant for the case of only one fermion type (or flavour), $N_f=1$. Consider now an arbitrary number of massless fermions $N_f\geqslant 2$. Then the multicomponent (in the sense of the flavour) fermionic field admits two independent infinitesimal transformations in the spirit of Eqs.~\eqref{gt} and \eqref{gt5},
\be
\left(\delta\psi^i\right)_V=-i\gt^a\left(t^a\right)_j^i\psi^j,\qquad
\left(\delta\psi^i\right)_A=-i\gt^{\prime a}\left(t^a\right)_j^i\gamma_5\psi^j,\qquad i,j=1..N_f,\qquad a=1..N_f^2-1,
\ee
where, as before, all $\gt$'s and $\gt'$'s are arbitrary constants and $t^a=\lambda^a/2$ (with $\{\lambda^a\}$ for $N_f^2-1$ Gell-Mann flavour matrices) are $N_f\times N_f$ traceless Hermitian matrices satisfying the $SU(N_f)$ algebra,
\be
[t^a,t^b]=if^{abc}t^c,
\ee
with $f^{abc}$ for the antisymmetric structure constants of the group. Then the vector and axial-vector currents,
\be
j_\mu^a=\bar{\psi}_i\gamma_\mu(t^a)^i_j\psi^j,\qquad
j_{5\mu}^a=\bar{\psi}_i\gamma_\mu\gamma_5(t^a)^i_j\psi^j,
\ee
are both conserved in the limit of massless fermions, and the corresponding charges,
\be
Q^a=\int d^3x j_0^a(x),\qquad Q_5^a=\int d^3x j_{50}^a(x),
\label{QQ5}
\ee
can be conventionally reshuffled as
\be
Q_R^a=\frac12(Q^a+Q_5^a),\qquad Q_L^a=\frac12(Q^a-Q_5^a),
\label{QLR}
\ee
with $Q_L^a$ and $Q_R^a$ generating two independent $SU(N_f)$ algebras,
\be
[Q_R^a,Q_R^b]=if^{abc} Q_R^c,\qquad
[Q_L^a,Q_L^b]=if^{abc} Q_L^c,\qquad
[Q_R^a,Q_L^b]=0.
\ee

Similarly to the $N_f=1$ case, even in the limit of vanishing fermion masses, divergence of the singlet axial-vector current is anomalous,
\be
\partial_\mu \Bigl[\bar{\psi}_i\gamma^\mu\gamma_5\psi^i\Bigr]=\frac{g^2N_f}{8\pi^2}\sum_{A=1}^{N_c^2-1}G_{\mu\nu}^A\tilde{G}^{\mu\nu;A},
\label{axveccurrconsNf}
\ee
with $g$ for the gauge coupling constant and $G_{\mu\nu}$ ($\tilde{G}_{\mu\nu}$) for the non-Abelian gauge field tensor (dual tensor) that transforms according to the adjoint representation of the colour gauge group.\footnote{A comprehensive review on the anomalies in gauge theories can be found, for example, in the book \cite{Bertlmann:1996xk}.} Therefore, for the number of fermion flavours $N_f\geqslant 2$, one has
\be
U(N_f)_L\otimes U(N_f)_R
\mathop{\longrightarrow}^{\rm Anomaly} U(1)_V\otimes SU(N_f)_L\otimes SU(N_f)_R,
\label{symschNf}
\ee
where, as before, the right arrow indicates an explicit breaking of the classical symmetry by the quantum anomaly in the divergence of the singlet axial-vector current in Eq.~\eqref{axveccurrconsNf}.\footnote{A typical scaling of the gauge coupling $g$ with the number of colours is $g\propto 1/\sqrt{N_c}$ \cite{tHooft:1973alw}, so the right-hand side in Eq.~\eqref{axveccurrconsNf} scales as ${\cal O}(1/N_c)$ and vanishes for $N_c\to\infty$.\label{foot:etapr}} The symmetry subgroup $SU(N_f)_L\otimes SU(N_f)_R$ in the scheme \eqref{symschNf} generated by $Q_R^a$ and $Q_L^a$ in Eq.~\eqref{QLR} is known as chiral symmetry that will be the main subject of the discussion in the following sections.

\section{Chiral symmetry in strong interactions}
\label{sec:QCD}

The symmetry scheme outlined above applies naturally to the theory of strong interactions known as QCD that also enters the SM together with the theories of electromagnetic and weak interactions. The fields of matter in this theory are quarks, and there are six different types of quarks conventionally referred to as quark flavours: ``$u$'' (``up''), ``$d$'' (``down''), ``$s$'' (``strange''), ``$c$'' (``charm''), ``$b$'' (``bottom''), and ``$t$'' (``top''). Noticeably, the hierarchy of their masses is rather peculiar, with the quark masses ranging from a few MeV for the lightest quarks to hundreds of GeV for the heaviest $t$ quark \cite{ParticleDataGroup:2024cfk},
\be
m_u\approx 2~\mbox{MeV},\quad m_d\approx 5~\mbox{MeV},\quad m_s\approx 94~\mbox{MeV},\quad m_c\approx 1.3~\mbox{GeV},\quad m_b\approx 4.2~\mbox{GeV},\quad m_t\approx 160..170~\mbox{GeV}.
\ee
The intrinsic scale of QCD, $\Lambda_{\rm QCD}\simeq 240~\mbox{MeV}$, allows one to separate the six quarks above into light ($u$, $d$, and $s$ with the masses below $\Lambda_{\rm QCD}$) and heavy ($c$, $b$, $t$ which are heavier than $\Lambda_{\rm QCD}$).
Notice that strong interactions are blind to the quark flavour and, in addition,
the mass difference $m_d-m_u$ can be disregarded in many applications in the physics of hadrons, that is strongly interacting composite objects made of quarks, their antiparticles (antiquarks), and gluons --- the quanta of the gauge field which mediate strong interactions between quarks. Then the number of light quark flavours in QCD is often set to $N_f=2$, and the corresponding approximate but rather accurate symmetry of strong interactions that emerges is know as isospin symmetry. If the difference between $m_q=(m_u+m_d)/2$ and the strange quark mass $m_s$ is also negligible in the studied problem, the approximate symmetry of the light quarks can be extended from isospin $SU(2)_f$ to flavour $SU(3)_f$. In this case, $N_f=3$. This symmetry is the cornerstone of the classification scheme suggested in 1961 by Gell-Mann \cite{Gell-Mann:1961omu} and Ne'eman \cite{Neeman:1961jhl} for the hadrons consisting of $u$, $d$, and $s$ quarks known at that time. The arising relations between the masses of such hadrons (the so-called Gell-Mann--Okubo mass formulae) \cite{Gell-Mann:1961omu,Okubo:1961jc} are fulfilled to a high precision. In many practical applications, an exact $SU(3)_f$ invariance is often assumed for the interaction (\emph{e.g.}, for coupling constants) while the mass difference $m_s-m_q$ is taken into account explicitly.
In this case, the number of light quark flavours is usually stated as $N_f=2+1$. Extension of chiral symmetry to $N_f=4$ and further does not have any practical sense given its strong explicit violation by the large masses of the heavy quarks.
On the contrary, since the masses of the light quarks are rather small in comparison with the typical scale of strong interactions, then setting $m_u=m_d=m_s=0$ is typically a good approximation known as the chiral limit of QCD since chiral symmetry $SU(3)_L\times SU(3)_R$ becomes exact in this limit.

\section{Different realisations of chiral symmetry}
\label{sec:CS}

\begin{figure}[t]
\centering
\begin{tabular}{ccccccc}
\includegraphics[width=0.18\textwidth]{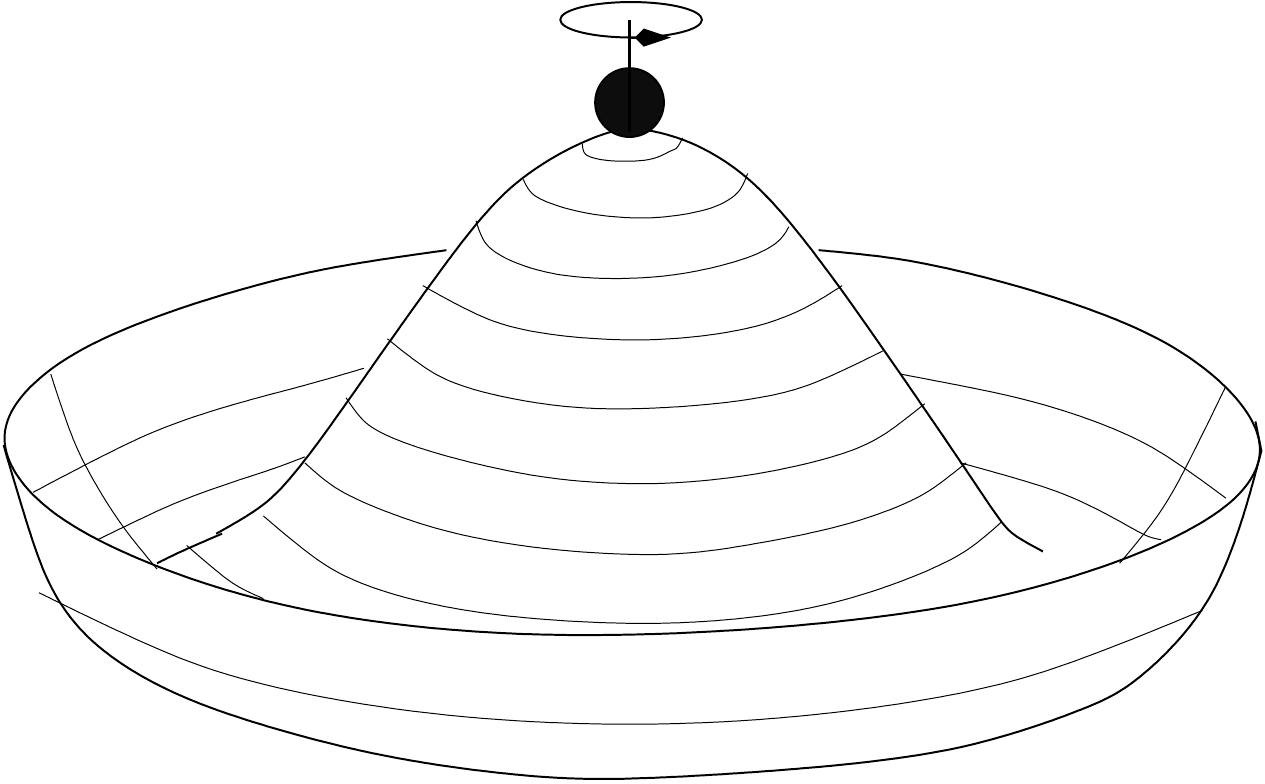}&&
\includegraphics[width=0.18\textwidth]{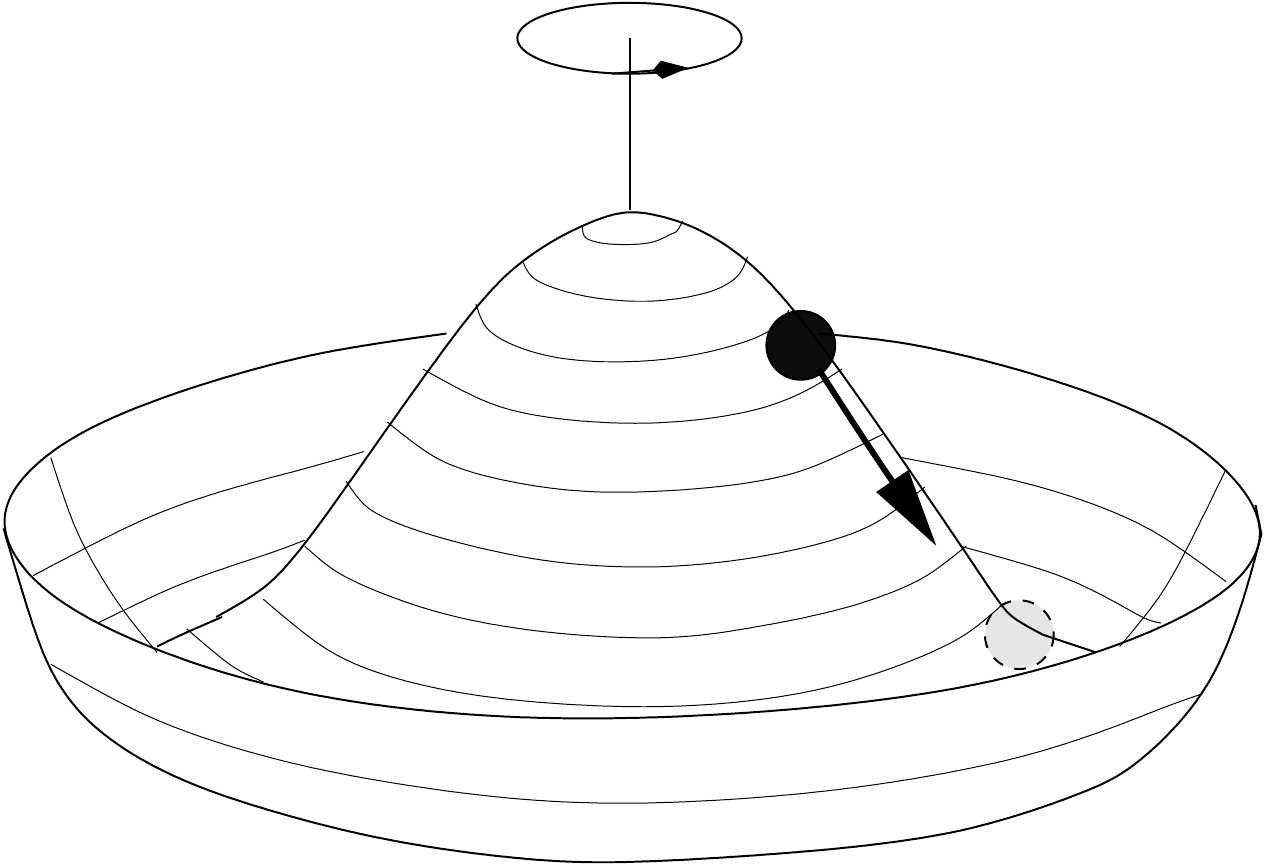}&&
\includegraphics[width=0.18\textwidth]{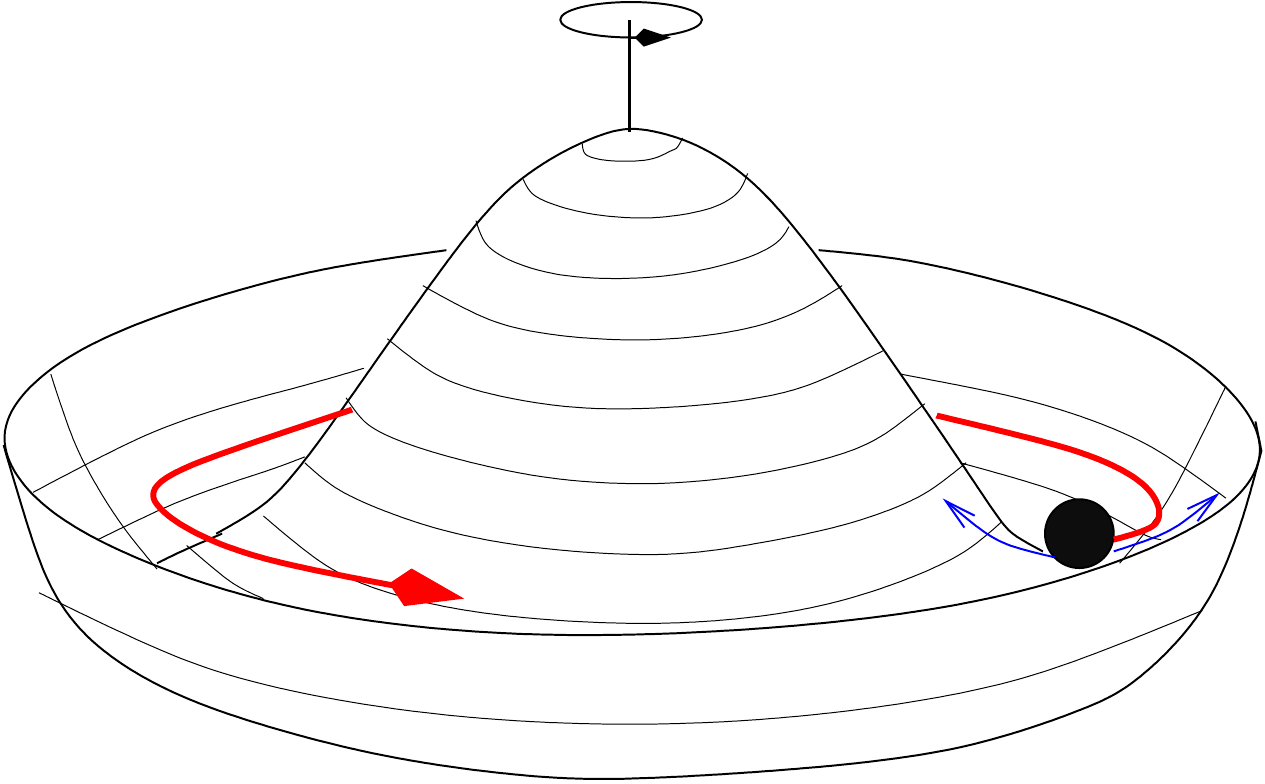}&&
\includegraphics[width=0.18\textwidth]{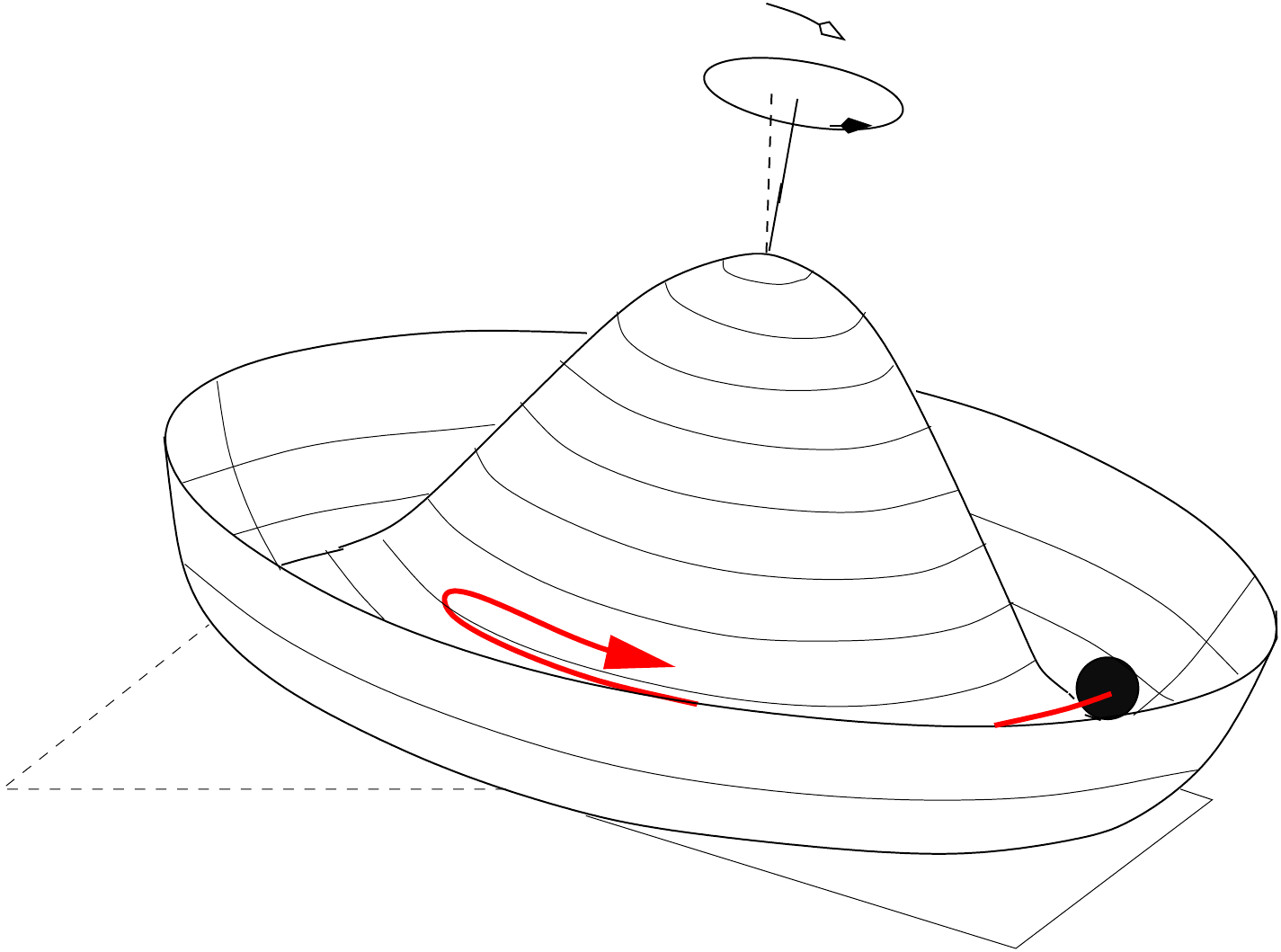}\\
(a)&&(b)&&(c)&&(d)
\end{tabular}
\caption{A simple mechanistic demonstration of the phenomenon of spontaneous symmetry breaking: (a) unstable but symmetric equilibrium point; (b) proceeding to stable but not symmetric equilibrium position; (c) ``massive'' (blue) and ``massless'' (red) excitations; (d) the effect of a slight explicit symmetry breaking. See the main text for further details and explanations.}
\label{fig:mex}
\end{figure}

Consider time derivatives of the charges in Eq.~\eqref{QQ5},
\be
\frac{\partial Q^a}{\partial t}=\int d^3x \frac{\partial j_0^a}{\partial t}=
-\int d^3x\; \nabla {\bm j}^a=-\oint_{S_\infty} d{\bm S}\;{\bm j}^a=0,\qquad
\frac{\partial Q_5^a}{\partial t}=\int d^3x \frac{\partial j_{50}^a}{\partial t}=
-\int d^3x\; \nabla {\bm j}_5^a=-\oint_{S_\infty} d{\bm S}\;{\bm j}_5^a=0,
\ee
where the currents conservation laws in Eqs.~\eqref{veccurrcons} and \eqref{axveccurrconsclass} were used as well as the assumption that all fields vanish at the infinitely remote three-dimensional surface $S_\infty$. Therefore, both charges $Q^a$ and $Q_5^a$ are constants of motions and as such commute with the Hamiltonian,
\be
[Q^a,H_{\rm QCD}]=[Q_5^a,H_{\rm QCD}]=0.
\label{QQ5H}
\ee
Na{\"i}vly, the latter observation is supposed to entail a relation between the masses $M_\pm$ of the opposite-parity hadronic states $\ket{h_\pm}$. Indeed, consider a positive-parity state $\ket{h_+}$ such that
\be
P\ket{h_+}=\ket{h_+},\qquad
H_{\rm QCD}\ket{h_+}=M_+\ket{h_+},
\ee
where $P$ is the parity transformation operator. An opposite-parity partner state $\ket{h_-}$ can be built by applying the axial charge to $\ket{h_+}$,
\be
\ket{h_-}=Q_5^a\ket{h_+},\qquad
P\ket{h_-}=-\ket{h_-},\qquad
H_{\rm QCD}\ket{h_-}=M_-\ket{h_-}.
\label{hmQ5hp}
\ee
On the other hand, since the axial charge commutes with the Hamiltonian, then
\be
H_{\rm QCD}\ket{h_-}=H_{\rm QCD}Q_5^a\ket{h_+}=
Q_5^aH_{\rm QCD}\ket{h_+}=M_+Q_5^a\ket{h_+}=M_+\ket{h_-},\quad\Longrightarrow\quad M_-=M_+.
\ee
If the above relation $M_-=M_+$ holds, the corresponding realisation of chiral symmetry is known as the Wigner--Weyl mode. In this mode, the opposite-parity states in the spectrum should be degenerate. However, the spectrum of the observed hadrons does not demonstrate this degeneracy. The easiest way to see it is to notice that the lightest pseudoscalar meson, the pion, does not have an equally light scalar partner and the lightest positive-parity baryon, the nucleon, does not have an equally light negative-parity partner.
It should be noted, however, that in the discussion above it was assumed that the lowest state of the system --- the vacuum $\ket{0}$ --- is chirally invariant, so $Q_5^a\ket{0}=0$.
However, this property does not necessarily hold since the vacuum state may not respect all the symmetries of the Hamiltonian of the theory.
To illustrate the pattern consider a simple mechanical system consisting of a little ball placed on the top of a Mexican hat as depicted in Fig.~\ref{fig:mex}(a). The hat shape provides a profile of the potential energy for the ball that possesses several remarkable features. On the one hand, the point on the top where the ball is placed corresponds to a maximum (in general, a local maximum is sufficient) of its potential energy. On the other hand, the potential is symmetric with respect to rotations around the vertical $z$ axis and the minimum of the energy is provided by a continuous circle of some fixed nonzero radius. If the ball does not experience any external impact, it can reside at rest on the top of the hat for an undetermined period of time, so its position can be regarded as a ``vacuum'' of the system. This vacuum is, however, unstable since it is energetically favourable for the ball to leave it and roll down from the hill to the valley --- see Fig.~\ref{fig:mex}(b). Then, given the energy loss because of the friction, the ball finally chooses a particular point of the equipotential circle to rest which will play a role of the new ``vacuum'' of the system. The rotational symmetry of the system is broken by this new vacuum. Note that the ball's moving along the equipotential circle does not change the energy of the system while pushing it in the radial direction increases the energy --- see Fig.~\ref{fig:mex}(c).
Translated into the language of quantum field theory (QFT), the
situation depicted in Fig.~\ref{fig:mex}(a)-(c) implies that
(i) under certain circumstances, the trivial (false) vacuum of the theory may be unstable;
(ii) the energetically favourable (true) vacuum may not hold the symmetry of the trivial vacuum although the Hamiltonian of the theory still possesses this symmetry --- this way the symmetry of the Hamiltonian is spontaneously broken by the vacuum;
(iii) spontaneous breaking of a continuous symmetry entails the appearance of a zero mode (massless particle after quantisation) in the spectrum of excitations (Goldstone boson), which is an essence of the Goldstone theorem \cite{Goldstone:1961eq}. Since in the real world the lightest $u$, $d$, and $s$ quarks are not strictly massless, chiral symmetry is slightly explicitly broken by their masses. For the simple mechanical system in Fig.~\ref{fig:mex} it implies that the Mexican hat is somewhat titled as depicted in Fig.~\ref{fig:mex}(d). Then, although there exists one favourable stable position for the ball that provides the minimum of the potential energy, the ball's moving along the circle increases the potential energy only marginally, so the corresponding excitation is not strictly massless but appears extremely light. For this reason, the corresponding nearly massless states in the spectrum of hadrons are often referred to as pseudo-Goldstone bosons.

The consideration above implies that there exists an alternative scenario of chiral symmetry realisation in QCD, known as the Nambu--Goldstone mode, in which the scheme in Eq.~\eqref{symschNf} further reads as
\be
U(N_f)_L\otimes U(N_f)_R
\mathop{\longrightarrow}^{\rm Anomaly} U(1)_V\otimes SU(N_f)_L\otimes SU(N_f)_R\mathop{\longrightarrow}^{\mbox{\tiny SBCS}} U(1)_V\otimes SU(N_f)_V.
\label{symschNf2}
\ee
Then a smoking gun for this scenario is the existence in the spectrum of excitations of a (nearly) massless pseudoscalar particle per each generator of the broken symmetry group (see, for example, the proof contained in \cite{Bernstein:1974rd}), referred to as $\pi$'s here, such that
\be
\braket{0|j_{5\mu}^a(x=0)|\pi^b(p)}=if_\pi p_\mu\delta^{ab},\quad a,b=1..N_f^2-1,
\label{sbcs}
\ee
with a nonvanishing constant $f_\pi$.\footnote{There exist two conventions for $f_\pi$ that differ by a factor $\sqrt{2}$ --- see Eq.~\eqref{GMOR} and footnote~\ref{foot:fpi} below. In the convention adopted here, $f_\pi\approx 92$~MeV.} For $N_f=2$, the number of such generators is 3, and indeed the spectrum of hadrons in QCD contains 3 nearly massless mesons: 2 charged and 1 neutral pion. For $N_f=3$, the 8 Goldstone bosons are the 3 pions ($\pi^0$, $\pi^\pm$), 4 kaons ($K^0$, $\bar{K}^0$, $K^\pm$), and 1 $\eta$-meson. Since $SU(3)_f$ symmetry is stronger explicitly broken by the strange quark mass, the kaons and $\eta$ appear to be heavier than the pions. However, all aforementioned pseudoscalar mesons would be massless in the strict chiral limit of $m_u=m_d=m_s=0$. The 9th pseudoscalar meson, $\eta'$, which naturally appears in the quark model scheme based on flavour symmetry $SU(3)_f$ \cite{Gell-Mann:1961omu,Neeman:1961jhl}, would remain massive even in the chiral limit since its mass is regulated by the axial anomaly in Eq.~\eqref{axveccurrconsNf}, which resolves the so-called $\eta$-$\eta'$ puzzle \cite{Weinberg:1975ui,Witten:1980sp,DiVecchia:1980yfw}. Meanwhile,  $\eta'$ would become massless too in the limit of an infinite number of colours --- see footnote \ref{foot:etapr}. Equation \eqref{sbcs} entails that
\be
\braket{0|\partial^\mu j_{5\mu}^a|\pi^b}=f_\pi M_\pi^2\delta^{ab},
\label{sbcs2}
\ee
which is considered to follow from an operator relation referred to as a partial conservation of the axial-vector current (PCAC) \cite{Adler:1964um,Weinberg:1966kf},
\be
\partial ^\mu j_{5\mu}^a=f_\pi M_\pi^2\pi^a.
\label{sbcs3}
\ee

\section{Exploring spontaneous breaking of chiral symmetry in quark models}
\label{sec:models}

To summarise the consideration in the previous section, since the experimentally observed spectrum of hadronic states in QCD does not demonstrate the degeneracy pattern inherent to the Wigner--Weyl realisation of chiral symmetry, we expect that it is realised according to the Nambu--Goldstone scenario. In this scenario, the physical vacuum of the theory is not chirally symmetric and thus the phenomenon of SBCS takes place,  which can be studied using various models for QCD. A pioneer quark model of this kind was suggested by Nambu and Jona-Lasinio (NJL model) in 1961 \cite{Nambu:1961tp,Nambu:1961fr}. For a comprehensive review and further developments of the model see, for example, \cite{Bernard:1987sg,Vogl:1991qt,Klevansky:1992qe,Hatsuda:1994pi,Buballa:2003qv,Volkov:2005kw}. Notice, that this model does not provide a systematic approach to QCD and is employed here for illustrative purposes only.

For two quark flavours, the Hamiltonian of the NJL model reads
\be
{\cal H}_{\rm NJL}=\int d^3 x\left\{
\psi^\dagger\left(-i\vec{\alpha}\cdot
{\bm\nabla}+\beta m_0
\right)\psi
-\frac{\lambda}{4}\left[(\bar \psi \psi)^2+(\bar \psi i\gamma_5\vec{\tau}\psi)^2\right]\right\},
\label{NJL}
\ee
with $\vec{\tau}$ for the isospin Pauli matrices, the coupling constant $\lambda$ defining the strength of the fermionic field self-interaction, and $m_0$ for the bare quark mass which will be set to zero below to approach the exact chiral limit. Although each individual contribution to the interaction term in Lagrangian \eqref{NJL} is not symmetric with respect to the chiral transformations from the $SU(2)_L\otimes SU(2)_R$ group, the relative coefficient is tuned in such a way that the sum of the two terms is chirally invariant.\footnote{For $N_f=1$ it is easy to find that $\left(\bar{\psi}\psi\right)^2+\left(\bar{\psi}i\gamma_5\psi\right)^2=
\left(\bar{\psi}_R\psi_L+\bar{\psi}_L\psi_R\right)^2-\left(\bar{\psi}_R\psi_L-\bar{\psi}_L\psi_R\right)^2
=4\left(\bar{\psi}_R\psi_L\right)\left(\bar{\psi}_L\psi_R\right)$, where chiral symmetry of the result is evident.}

The mass operator that arises due to the fermionic field self-interaction reads
\be
i\Sigma=2\includegraphics[width=1cm]{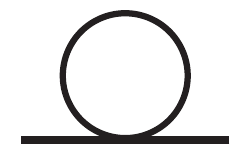}=
i\lambda N_c\int\frac{d^4p}{(2\pi)^4}\mbox{Tr}\;\Bigl[S(p)\Bigr]=2i\lambda m N_c \int\frac{d^3p}{(2\pi)^3}\frac{1}{\sqrt{\vep^2+m^2}},
\label{Sigma}
\ee
where the traces over the colour and flavour matrices were taken explicitly, so Tr stands for the trace from the Dirac matrices, and $S(p)$ is the quark Green function.  Equation~\eqref{Sigma} is a self-consistency condition for the physical fermion mass $m=m_0+\Sigma$ which, for $m_0=0$, reads
\be
m\left(\frac12-\lambda N_c \int\frac{d^3p}{(2\pi)^3}\frac{1}{\sqrt{\vep^2+m^2}}\right)=0.
\label{mg1}
\ee
Notice that the integral in parentheses diverges quadratically and needs to be regularised, for example, by cutting the loop momentum from above as $|\vep|<\Lambda$. After that, two essentially different regimes of the model can be identified. In the weak coupling regime of $\lambda<\lambda_{\rm crit}=2\pi^2/(N_c\Lambda^2)$, the expression in parentheses never hits zero, so equation \eqref{mg1}
possesses only a trivial solution of $m=0$. On the contrary, in the strong coupling regime of $\lambda>\lambda_{\rm crit}$, equation \eqref{mg1} has a nontrivial solution with $m\neq 0$. Then the vacuum expectation value of a chirally noninvariant operator $\bar{\psi}\psi$ (see Eq.~\eqref{vecscalRL} and the discussion below it), known as the chiral condensate, can be found to take a nonvanishing value in this phase of the model,
\be
\braket{\bar{\psi}\psi}=-2N_c\int\frac{d^4p}{(2\pi)^4}\mbox{Tr}\;\Bigl[S(p)\Bigr]=
-4mN_c\int\frac{d^3p}{(2\pi)^3}\frac{1}{\sqrt{\vep^2+m^2}}
=-\frac{2m}{\lambda}\neq 0.
\ee
Therefore, we started from a chirally symmetric theory with massless fermions and a gapless spectrum of excitations and arrived at massive physical excitations and a gap in the spectrum. For this reason, Eq.~\eqref{mg1} is often referred to as gap or mass-gap equation.

As was explained above, to unambiguously establish SBCS in the vacuum in general case, one is to ensure the appearance of $N_f^2-1$ (nearly) massless pseudoscalar Goldstone bosons which possess the property \eqref{sbcs}.
At the same time, these bosons need to support an interpretation as the lightest quark-antiquark states (mesons) in the spectrum of hadrons. This dual nature of the Goldstone bosons can be understood employing a suitable quark model. It has to be noticed, however, that the NJL model  has several shortcomings that foil employing it as a ``microscopic'' investigation tool for SBCS. These shortcomings include (i) the absence of confinement that is an intrinsic property of QCD preventing coloured objects such as quarks from existing as free particles; (ii)
the absence of a natural scale, so all physical quantities acquire dimension from the regulator (for example, a sharp momentum cut-off $\Lambda$ employed above);
(iii) a vague relation between the model and full QCD where the interaction between quarks is mediated by vector gluons.
These shortcomings can be cured by proceeding from the NJL model described by the  Hamiltonian in Eq.~\eqref{NJL} to a generalised Nambu--Jona-Lasinio (GNJL) model with the Hamiltonian
\be
{\cal H}_{\rm GNJL}=\int d^3x\left\{\psi^\dagger\left(-i\vec{\alpha}\cdot
{\bm\nabla}+\beta m_0\right)\psi-\frac{1}{2} \int d^3y\;\rho^A(\vex)K_{AB}(|\vex-\vey|)\rho^B(\vey)\right\},
\label{GNJL}
\ee
which incorporates the interaction of two quark colour charge densities\footnote{More generally, quark currents $j_\mu^A=\bar{\psi}\gamma_\mu\frac{\lambda_c^A}{2}\psi$; however, considering only their temporal parts, $\rho^A=j^A_0$, makes the calculations easier without changing the argument.},
$\rho^A=\psi^\dag\frac{\lambda_c^A}{2}\psi$ ($A=1..N_c^2-1$, with $N_c$ for the number of fundamental colours)\footnote{We use the subscript $c$ in $\lambda_c^A/2$ to tell the generators of the fundamental representation of the colour $SU(3)$ group from the flavour matrices used above.}, taken at different spatial points $\vex$ and $\vey$, via an instantaneous\footnote{The relative time $x_0-y_0$ is irrelevant for the model that admits, therefore, a description in terms of the Hamiltonian responsible for the evolution in the time $t=x_0=y_0$.} confining kernel,
\be
K_{AB}(|\vex-\vey|)=\delta_{AB}V_0(|\vex-\vey|).
\label{Kab}
\ee
For simplicity, we consider the case of $N_f=1$ while generalisation to a larger number of quark flavours is straightforward. Notice that the model lacks dynamical gluons, so there is no axial anomaly and chiral symmetry $U(1)_L\otimes U(1)_R$ is broken spontaneously. Then the corresponding Goldstone boson that appears in the result of SBCS will be referred to as the chiral pion. Although this model does not provide a systematic approach to QCD either, it
allows one to investigate the phenomenon of SBCS in some detail and as such has a rich history and is widely discussed in the literature --- see, for example, \cite{Amer:1983qa,LeYaouanc:1983it,LeYaouanc:1983huv,LeYaouanc:1984ntu,Adler:1984ri,Kocic:1985uq,Bicudo:1989sh,Bicudo:1989si,Bicudo:2002eu,Llanes-Estrada:1999nat,Bicudo:2003cy,Nefediev:2004by,Alkofer:2005ug,Wagenbrunn:2007ie,Bicudo:2008kc} and references therein. An explicit form of the confining potential $V_0(r)$ does not change the qualitative argument but it is crucial that the kernel in Eq.~\eqref{Kab} introduces a physical scale with the dimension of mass. This property is evident for the most popular choice for $V_0(r)$ as a power-like function of the quarks separation,
\be
V_0(r)=K_0^{\power+1}r^\power,
\label{Vconf}
\ee
with $K_0$ for the above physical scale. In general, the range $-1<\power<3$ was addressed in the literature, with $\power\to-1$ approaching the parameter-free colour Coulomb potential and $\power\geqslant 3$ (more generally, $\power$ exceeding the number of spatial dimensions) resulting in intractable infrared singularities. The limit of $\power\to 0$ implies an appropriate modification of the potential and
results in a logarithmic interaction,
\be
V_0(r)\to \tilde{V}_0(r)=\lim_{\power\to 0}\frac{K_0}{\power}\Bigl[(K_0r)^\power-1\Bigr]=K_0\log(K_0r),
\ee
and the harmonic oscillator potential with $\power=2$ is famous for leading to second-order differential equations rather than integral equations.  Meanwhile, the most phenomenologically adequate choice of the confining interaction is given by the linear function with $\power=1$. In this case, the GNJL model can be viewed as a four-dimensional generalisation of the well-known 't~Hooft model for QCD in two dimensions \cite{tHooft:1974pnl,Bars:1977ud,Li:1986gf,Kalashnikova:2001df,Glozman:2012ev} or as an approximation for Coulomb-gauge QCD \cite{Christ:1980ku,Szczepaniak:2001rg,Feuchter:2004mk,Reinhardt:2017pyr,Nguyen:2024ikq} if only the Coulombic confining part of the interaction is retained. A study of the model is performed with the large-$N_c$ logic in mind. Then $V(r)=C_F V_0(r)$, with
$C_F=\mbox{Tr}[(\lambda_c^A/2)\cdot(\lambda_c^A/2)]=(N_c^2-1)(2N_c)$
for the eigenvalue of the fundamental Casimir operator, should reach a finite limit as $N_c\to\infty$. For a power-like potential in Eq.~\eqref{Vconf} it implies that, for large $N_c$, $K_0$ scales as $N_c^{1/(\power+1)}$.
To simplify notations, in what follows the colour indices will be suppressed altogether since they can be trivially restored. Note that summation over colours is always implied for the products of quark and antiquark creation and annihilation operators.
Also, as before, we consider an exact chiral limit and set $m_0=0$.
The discussion below mainly follows the lines of \cite{Bicudo:1989sh,Bicudo:1989si,Nefediev:2004by}.

\begin{figure}[t]
\centering
\includegraphics[width=0.35\textwidth]{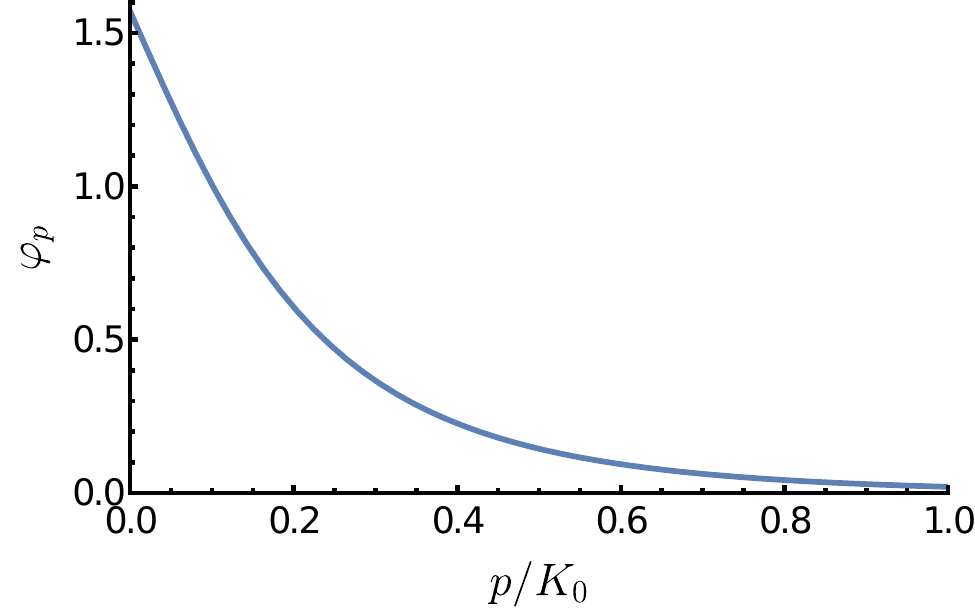}
\caption{A typical profile of the chiral angle --- solution to the mass-gap equation in Eq.~\eqref{massgap}.}
\label{fig:vp}
\end{figure}

The Hamiltonian in Eq.~\eqref{GNJL} describes self-interaction of the fermionic field through the confining kernel \eqref{Kab}, so the physical, ``dressed'', degrees of freedom in the system differ from the ``bare'' quarks that experience no interaction. This effect is best parametrised through the so-called chiral angle $\vp_p$ --- a function of the momentum $p=|\vep|$ that by convention varies in the range
$-\frac{\pi}{2}<\vp_p\leqslant \frac{\pi}{2}$ with the boundary
conditions $\vp(0)=\frac{\pi}{2}$ and $\vp_p(p\to\infty)\to 0$. In terms of the chiral angle the fermionic bispinor amplitudes can be presented in the form
\be
u_s(\vep)=\ds\frac{1}{\sqrt{2}}\left[\sqrt{1+\sin\vp_p}+
\sqrt{1-\sin\vp_p}\;(\vec{\alpha}\hat{\vep})\right]u_s(0),\qquad
v_{-s}(-\vep)=\ds\frac{1}{\sqrt{2}}\left[\sqrt{1+\sin\vp_p}-
\sqrt{1-\sin\vp_p}\;(\vec{\alpha}\hat{\vep})\right]v_{-s}(0),
\label{uandv}
\ee
for the quark and antiquark, respectively, where the rest-frame bispinors are
\be
u_s(0)=
\begin{pmatrix}
w_s \\ 0
\end{pmatrix}
,\qquad
v_{-s}(0)=-i\gamma^2 u_s^*(0)=
\begin{pmatrix}
0\\
i\sigma_2 w_s^*
\end{pmatrix},
\label{spinors}
\end{equation}
with $\gamma^2(\sigma_2)$ for the second Dirac(Pauli) matrix, $s=\pm 1/2$ for the helicity, and $(w_s)_{s'}=\delta_{ss'}$ for the rest-frame spinor. Then the spectral decomposition for the fermion field reads
\be
\psi(\vex,t)=\sum_{s=\pm1/2}\int\frac{d^3p}{(2\pi)
^3}e^{i\vep\vex}\Bigl[e^{-iE_pt}b_s(\vep)u_s(\vep)\\
+e^{iE_pt}d_s^\dagger(-\vep)v_s(-\vep)\Bigr],
\label{psi}
\ee
where $b$ and $d$ are the quark and antiquark annihilation operators, respectively, and $E_p$ is the dispersion law of the dressed fermion that is also parametrised in terms of the chiral angle.
The Hamiltonian of the model in Eq.~\eqref{GNJL} can be normally ordered in terms of the quark/antiaquark creation and annihilation operators to acquire a general structure
\be
{\cal H}^{\rm GNJL}=H_0\;+:H_2:+:H_4:,
\label{H024}
\ee
where the subscript 0, 2, or 4 indicates the power of the creation and annihilation operators contained in the respective term.

The trivial vacuum of the model, $\ket{0}_0$, corresponds to no quark dressing ($\vp_p\equiv 0$) and is annihilated by the bare fermion operators $b^{(0)}$ and $d^{(0)}$.
This vacuum is chirally symmetric but does not provide the minimum of the vacuum energy --- to ensure the latter the vacuum energy, that sums the contributions from zero-point oscillations, needs to be minimised with respect to $\vp_p$,
\be
\frac{\delta}{\delta\vp_p}\braket{0|{\cal H}^{GNJL}|0}=\frac{\delta H_0[\vp_p]}{\delta\vp_p}=0,\qquad H_0=\sum_{\rm colour}\sum_{\rm spins}\sum_p\left(-\frac12 E_p\right)=-N_cV \int\frac{d^3p}{(2\pi)^3}E_p,
\label{massgap}
\ee
where $\ket{0}\neq\ket{0}_0$ is the physical vacuum and $V$ is the 3D volume of space. Equation \eqref{massgap} for a function of the momentum $\vp_p$ is the mass-gap equation in the GNJL model which, therefore, is a more sensitive investigation tool for SBCS than the NJL model. Alternatively, the mass-gap equation can be obtained from the Dyson series for the dressed quark propagator considered in the rainbow approximation, which amounts to summing only planar diagrams most relevant in the large-$N_c$ limit. The profile of a typical solution to Eq.~\eqref{massgap} is shown in Fig.~\ref{fig:vp}. One can conclude from this figure that, quite naturally, the dressing effect is strongest for small momenta, where $\vp_p$ deviates from zero. Then the physical vacuum $\ket{0}$ that provides a minimum of the vacuum energy is
\begin{equation}
|0\rangle=S_0\ket{0}_0=e^{Q^\dagger-Q}|0\rangle_0,\qquad Q^\dagger=\frac12\sum_{p}\vp_p C_p^\dagger,\qquad
C_p^\dagger=\sum_{s,s'=\pm1/2}b^{(0)\dagger}_s(\vep)\Bigl[({\bm \sigma}\hat{{\bm p}})i\sigma_2\Bigr]_{ss'}d^{(0)\dagger}_{s'}(\vep),
\label{S00}
\end{equation}
where the three-vector $\vesig$ is formed by the Pauli matrices and the operator
$C_p^\dag$ creates a quark-antiquark pair with the relative momentum $\vep$ and the quantum numbers of the vacuum $0^{++}$ ($^3P_0$ pair). The new vacuum $\ket{0}$ is annihilated by the dressed quark and antiquark operators $b=S_0b^{(0)}S_0^\dag$ and $d=S_0d^{(0)}S_0^\dag$. Expanding the exponent in $S_0$ and taking into account the Pauli exclusion principle for the quarks and antiquarks, one can straightforwardly arrive at a simple representation for the physical vacuum,
\begin{equation}
|0\rangle=\mathop{\prod}\limits_{p}\left[\sqrt{w_{0p}}+
\frac{1}{\sqrt{2}}\sqrt{w_{1p}}C^\dagger_p
+\frac12\sqrt{w_{2p}}C^{\dagger 2}_p\right]|0\rangle_0,\quad\mbox{with}\quad
w_{0p}=\cos^4\frac{\vp_p}{2},~
w_{1p}=2\sin^2\frac{\vp_p}{2}\cos^2\frac{\vp_p}{2},~
w_{2p}=\sin^4\frac{\vp_p}{2}.
\label{nv}
\end{equation}
The coefficients satisfy a relation $w_{0p}+w_{1p}+w_{2p}=1$ and
admit an interpretation as the probabilities of having no, one, and two $q\bar{q}$ pairs with the given relative momentum $p$. The physical vacuum is normalised to unity and orthogonal to the trivial one,
\be
\langle 0|0\rangle=\prod_p(w_{0p}+w_{1p}+w_{2p})=1,\qquad
\braket{0|0}_0=\exp\left[\sum_p \log\left(\cos^2\frac{\vp_p}{2}\right)\right]
=\exp\left[V\int\frac{d^3p}{(2\pi)^3}\ln\left(\cos^2\frac{\vp_p}{2}\right)\right]
\mathop{\longrightarrow}\limits_{V\to \infty}0.
\ee

Note that the part $:H_2:$ of the Hamiltonian in Eq.~\eqref{H024} that is quadratic in the quark creation and annihilation operators takes a diagonal form in the basis of the dressed operators,
\be
:H_2:=\sum_{s=\pm1/2}\int
\frac{d^3 p}{(2\pi)^3} E_p\Bigl[b^\dagger_s({\bm p}) b_s({\bm p})+d^\dagger_s(-{\bm p}) d_s(-{\bm p})\Bigr],
\label{H2diag}
\ee
so, alternatively, the same mass-gap equation could be derived from a fermionic Bogoliubov transformation
applied to the bare quarks. In terms of the dressed-quark operators, the quartic term $:H_4:$ in Eq.~\eqref{H024} scales as $1/\sqrt{N_c}$ and is suppressed in the limit of $N_c\to\infty$.

The picture of the physical vacuum as a medium filled with condensed composite objects made of fermion Cooper-like pairs (therefore, bosons!) resembles the Bardeen, Cooper, and Schrieffer (BCS) theory of superconductivity. For this reason, the physical vacuum $\ket{0}$ is often called the BCS vacuum. The form of the pair creation operator $Q^\dagger$ in Eq.~\eqref{S00} suggests that the chiral angle $\vp_p$ can also be interpreted as the wave function of the scalar quark-antiquark pairs condensed in this vacuum. The chiral condensate in the physical vacuum is calculated as
\be
\braket{\bar{\psi}\psi}=-\frac{N_c}{\pi^2}\int_0^{\infty}dp\;p^2\sin\vp_p\neq 0,
\label{chircond}
\ee
signalling that chiral symmetry is spontaneously broken. To ensure it is indeed the case let us check the structure of the axial charge operator defined in Eq.~\eqref{QQ5} and, for $N_f=1$, straightforwardly expressed in terms of the dressed quark operators as
\be
Q_5=
\sum_{s,s'=\pm1/2}
\int\frac{d^3p}{(2\pi)^3}
\Bigl[\cos\vp_p({\bm\sigma} \hat{\vep})_{ss'}\left(b_s^\dagger(\vep) b_{s'}^j(\vep)
+d_s^\dagger(-\vep) d_{s'}^j(-\vep)\right)
+\sin\vp_p(i\sigma_2)_{ss'}\left(b_s^\dagger(\vep) d^{\dagger}_{s'}(-\vep)+d_s(-\vep)b_{s'}^j(\vep)\right)\Bigr].
\label{ChiralChar}
\ee
The first term on the right-hand side is diagonal in the creation and annihilation operators and contains a $P$-odd operator $(\vesig\hat{\vep})$, so it flips the parity of the hadronic state it is applied to. As discussed in Sec.~\ref{sec:CS}, it is a typical behaviour of the axial charge operator in the Wigner--Weyl mode, so it should not come as a surprise that this term is maximal when $\vp_p=0$. On the contrary, the second contribution to the axial charge in Eq.~\eqref{ChiralChar} is anomalous (contains $b^\dagger d^\dagger$ and $db$), so it creates and annihilates quark-antiquark pairs with the quantum numbers $0^{-+}$ as provided by the operator $i\sigma_2$. In other words, as required by the necessary and sufficient condition of SBCS in Eq.~\eqref{sbcs}, the axial charge operator has a nonvanishing matrix element between the vacuum and a pseudoscalar state that can be identified as the Goldstone boson of SBCS. Naturally, the momentum distribution for this state, specified by the function $\sin\vp_p$, vanishes in the trivial vacuum with $\vp_p=0$.

\section{The chiral pion}
\label{sec:pion}

In the previous section, it was demonstrated how the effect of SBCS takes place in the GNJL model and, in accordance with general principles, there appears a Goldstone boson associated with it. On the other hand, the GNJL model describes confined quarks, so this boson must be nothing else but the lightest pseudoscalar meson in the spectrum of hadrons --- the pion.
To see it explicitly we proceed beyond the BCS level and take into account the part of the Hamiltonian $:H_4:$
that describes the interaction between dressed quarks. To this end the entire model needs to be reformulated in terms of the operators creating and annihilating colourless quark-antiquark pairs \cite{Nefediev:2004by,Kalashnikova:2017ssy},
\be
M_{ss'}^\dagger(p,p')=\frac{1}{\sqrt{N_c}} b_s^\dagger(p')d_{s'}^\dagger(-p),\qquad
M_{ss'}(p,p')=\frac{1}{\sqrt{N_c}} d_s(-p)b_{s'}(p').
\label{MMdag}
\ee
Then, up to $N_c$-suppressed terms, the Hamiltonian \eqref{GNJL} takes the form, schematically (momentum dependence and integrals are skipped),
\be
H\sim H_0+M^\dagger M+\frac12\left(M^\dagger M^\dagger +MM\right)
\label{HMM}
\ee
and, therefore, is subject to a bosonic Bogoliubov transformation (also schematically)
\be
m^\dagger\sim \vp^+ M^\dagger+\vp^- M,\qquad m\sim \vp^+M+\vp^-M^\dagger,\qquad (\vp^+)^2-(\vp^-)^2=1,
\label{mmm}
\ee
where the operators $m^\dagger$ and $m$ create and annihilate physical quark-antiquark mesons. The Bogoliubov amplitudes $\vp^+$ and $\vp^-$ should be interpreted then as two wave functions of a given meson describing the forward and backward in time motion of the quark-antiquark pair, respectively. It makes an essential difference between the field-theory-driven GNJL quark model \eqref{GNJL} and simpler potential quark models based on quantum mechanics --- in the latter, the backward in time motion is neglected altogether, which makes it impossible for such models to capture the chiral physics. The instantaneous nature of the interquark interaction through the confining kernel in Eq.~\eqref{Kab} requires that the quark and antiquark inside of a meson always proceed from forward to backward in time motion and vice verse \emph{in unison}. The requirement that the Hamiltonian in Eq.~\eqref{HMM} should take a diagonal form in terms of the operators $m^\dagger$ and $m$ results in a system of two coupled equations for the two wave functions $\vp^\pm$ rather than a single equation for one wave function.\footnote{For this reason, in the literature, the corresponding approach is often referred to as an energy-spin formalism.} Alternatively, these equations follow from a Bethe--Salpeter equation for the matrix meson--quark--antiquark vertex function derived in the ladder approximation, which amounts to summing only planar diagrams in the large-$N_c$ limit (see \cite{Llewellyn-Smith:1969bcu} for various details of the Bethe--Salpeter formalism for mesons). We refrain from quoting these equations in the general form (see, for example, \cite{Kalashnikova:2017ssy} and references therein for details) and only do it for the pion,
\begin{align}
[2E_p-M_\pi]\vp_\pi^+(p)=\ds\int\frac{\ds q^2dq}{\ds (2\pi)^3}
[T^{++}_\pi(p,q)\vp_\pi^+(q)+T^{+-}_\pi(p,q)\vp_\pi^-(q)],\nonumber\\[-3mm]
\label{bsp}\\[-3mm]
[2E_p+M_\pi]\vp_\pi^-(p)=\ds\int\frac{\ds q^2dq}{\ds (2\pi)^3}
[T^{-+}_\pi(p,q)\vp_\pi^+(q)+T^{--}_\pi(p,q)\vp_\pi^-(q)],\nonumber
\end{align}
where the strict chiral limit of $m_0=0$ is slightly relaxed, so the pion has a little mass $M_\pi$, and
\begin{align}
&T_\pi^{++}(p,q)=T_\pi^{--}(p,q)=-\int d\Omega_q V({\bm
p}-{\bm q})
\left[\cos^2\frac{\vp_p-\vp_q}{2}-\frac{1-(\hat{{\bm p}}\cdot\hat{{\bm q}})}{2}\cos\vp_p\cos\vp_q\right]
\equiv T^{\rm diag}_\pi
,\nonumber\\[-2mm]
\label{pia}\\[-2mm]
&T_\pi^{+-}(p,q)=T_\pi^{-+}(p,q)=-\int d\Omega_q V({\bm
p}-{\bm q}) \left[\sin^2\frac{\vp_p-\vp_q}{2}+\frac{1-(\hat{{\bm
p}}\cdot\hat{{\bm q}})}{2}\cos\vp_p\cos\vp_q\right]\equiv
T^{\rm off-diag}_\pi.
\nonumber
\end{align}
In the vicinity of the chiral limit ($M_\pi\to 0$), one can find for the normalised pion wave functions:
\be
\vp_\pi^\pm(p)=\frac{2\sqrt{\pi
N_c}}{f_\pi}\left[\sin\vp_p \pm
M_\pi\Delta_p+\ldots\right],\qquad
\int\frac{p^2dp}{(2\pi)^3}\left[\vp_\pi^{+2}(p)-\vp_\pi^{-2}(p)\right]=2M_\pi,
\label{vppm}
\ee
where the ellipsis denotes neglected terms of higher orders in the pion mass, the correction function $\Delta_p$ is a solution of the equation
\be
2E_p\Delta_p=\sin\vp_p+\int\frac{d^3q}{(2\pi)^3}V(\vep-\veq)\Bigl(\sin\vp_p\sin\vp_q+(\hat{\vep}\cdot\hat{\veq}
)\cos\vp_p\cos\vp_q\Bigr)\Delta_q,
\ee
and the pion decay constant is calculated as
\be
f_\pi^2=\frac{N_C}{\pi^2}\int_0^\infty p^2dp\Delta_p\sin\vp_p.
\label{fpidef}
\ee
Then the famous Gell-Mann--Oakes--Renner relation \cite{GellMann:1968rz} holds, with the chiral condensate defined in Eq.~\eqref{chircond}.\footnote{If $\braket{\bar{\psi}\psi}$ on the right-hand side is as a sum of the contributions from all light-quark flavours then the numerical value of $f_\pi^2$ on the left-hand side re-scales accordingly.\label{foot:fpi}}
\be
f_\pi^2M_\pi^2=-2m_0\braket{\bar{\psi}\psi}.
\label{GMOR}
\ee

Let us now proceed to the exact chiral limit of $M_\pi=0$ to see that the bound state equation for the pion in Eq.~\eqref{bsp} can be formulated as a single equation for its wave function (up to the normalisation factor) $\vp_\pi(p)\equiv\vp_\pi^+(p)=\vp_\pi^-(p)=\sin\vp_p$,
\be
2E_p\vp_\pi(p)=\int\frac{q^2dq}{(2\pi)^3}[T_\pi^{\rm diag}(p,q)+T_\pi^{\rm off-diag}(p,q)]\vp_\pi(q)=
-\int\frac{d^3q}{(2\pi)^3}V({\bm p}-{\bm q})\vp_\pi(q),
\ee
or, equivalently, in coordinate space,
\be
[2E_p+V(r)]\vp_\pi=0.
\label{bspion}
\ee
This relation can be readily recognised as a centre-of-mass Salpeter-like equation for a quark and antiquark, each with the ``kinetic energy'' $E_p$, bound by the confining potential $V(r)$ to a massless pion. It is important to notice, however, that the energy of the dressed quark $E_p$ essentially departs from the free particle dispersive law $E_p^{(0)}=\sqrt{\vep^2+m^2}$ and, what is crucial, becomes negative at small momenta to compensate for the purely positive contribution from the confining potential in Eq.~\eqref{bspion}. The latter observation explains how a massless pion can arise in a quark model. Notice that, in the exceptional case of the chiral pion, the backward in time motion of the quarks in it plays as essential role as their forward in time motion. For this reason, it cannot be neglected without violating fundamental properties of the system.
Last but not least, it can be verified that the pion bound-state equation in Eq.~\eqref{bspion} is equivalent to the mass-gap equation previously derived at the BCS level of dressed ``free'' quarks in Eq.~\eqref{massgap}. This way the same chiral pion indeed appears in the theory twice: as the Goldstone boson of SBCS and as the lightest pseudoscalar meson in the spectrum of hadrons. To see how the specific properties of the pions arise in different approaches to QCD such as, for example, the functional methods, check the respective chapter \cite{Eichmann:2025wgs}.

We can now revisit the axial charge operator in Eq.~\eqref{ChiralChar}
to see that, as per Eq.~\eqref{vppm},
the second term on the right-hand side indeed creates or annihilates
a pion in its rest frame, so $\braket{0|Q_5|\pi(\vep=0)}=if_\pi M_\pi$,\footnote{We disregard the $N_c$-suppressed difference between the vacuum annihilated by the meson operators $m$ in Eq.~\eqref{mmm} and the BCS vacuum $\ket{0}$.} in agreement with the criterion for SBCS in Eq.~\eqref{sbcs}. Then the action of the axial charge on a hadronic state is twofold: It flips the parity of the state or creates a pion, which explains why the relation between $\ket{h_\pm}$ in Eq.~\eqref{hmQ5hp} fails in the Nambu--Goldstone mode.

\section{Conclusions}
\label{sec:conc}

Chiral symmetry plays a key role in understanding the dynamics of the SM and is a cornerstone in the theoretical framework of strong interactions, providing deep insights into the fundamental structure of QCD. Interpretation of the spectrum of hadrons, its structure, degeneracy, and
properties of strong interactions at
low-energies could not be possible without a comprehensive insight into the nonperturbative physics related to chiral symmetry and its spontaneous breaking in the vacuum of QCD. Given quite unique properties and a very low mass of the Goldstone bosons related to SBCS, they play a distinguished role in strong interactions, especially in the low-energy limit where they represent the most relevant degrees of freedom in the effective low-energy theory for QCD known as the Chiral Perturbation Theory --- for further details see the respective chapter \cite{Meissner:2024ona} as well as its extended version in \cite{Meissner:2022cbi} and the references therein.
Restoration of chiral symmetry at high temperatures and baryon densities is a key area of research, relevant for heavy-ion collisions and studies of the early universe and neutron stars. Thus future research, particularly in high-energy experiments and lattice QCD simulations, should further illuminate the role of chiral symmetry in the dynamics of strong interaction and its broader implications in particle physics and cosmology.

\begin{ack}[Acknowledgments]%
I express my gratitude to my colleagues and collaborators for
enlightening me on the physics relevant to this work. I am especially indebted to (in alphabetic order) Pedro Bicudo, Leonid Glozman, Christoph Hanhart, Yulia Kalashnikova, Emilio Ribeiro, Yury Simonov, and many others who shared with me their understanding of chiral physics.
I would also like to thank Christoph Hanhart and Ulf Mei{\ss}ner for reading this text and providing their critical comments as well as Andreas Wirzba for the pictures files illustrating the phenomenon of SBCS.
This work was supported by Deutsche Forschungsgemeinschaft (Project No. 525056915).
\end{ack}

\begin{thebibliography*}{10}
\providecommand{\bibtype}[1]{}
\providecommand{\url}[1]{{\tt #1}}
\providecommand{\urlprefix}{URL }
\expandafter\ifx\csname urlstyle\endcsname\relax
  \providecommand{\doi}[1]{doi:\discretionary{}{}{}#1}\else
  \providecommand{\doi}{doi:\discretionary{}{}{}\begingroup
  \urlstyle{rm}\Url}\fi
\providecommand{\bibinfo}[2]{#2}
\providecommand{\eprint}[2][]{\url{#2}}
\makeatletter\def\@biblabel#1{\bibinfo{label}{[#1]}}\makeatother

\bibtype{Book}%
\bibitem{Coleman:1985rnk}
\bibinfo{author}{Sidney Coleman}, \bibinfo{title}{{Aspects of Symmetry}:
  {Selected Erice Lectures}}, \bibinfo{publisher}{Cambridge University Press},
  \bibinfo{address}{Cambridge, U.K.} \bibinfo{year}{1985}, ISBN
  \bibinfo{isbn}{978-0-521-31827-3},
  \bibinfo{doi}{\doi{10.1017/CBO9780511565045}}.

\bibtype{Book}%
\bibitem{Georgi:1999wka}
\bibinfo{author}{Howard Georgi}, \bibinfo{title}{{Lie algebras in particle
  physics}}, \bibinfo{comment}{vol.} \bibinfo{volume}{54},
  \bibinfo{edition}{second} ed., \bibinfo{publisher}{Perseus Books},
  \bibinfo{address}{Reading, MA} \bibinfo{year}{1999}.

\bibtype{Book}%
\bibitem{Georgi:2000vve}
\bibinfo{author}{Howard Georgi}, \bibinfo{title}{{Lie Algebras In Particle
  Physics : from Isospin To Unified Theories}}, \bibinfo{publisher}{Taylor \&
  Francis}, \bibinfo{address}{Boca Raton} \bibinfo{year}{2000}, ISBN
  \bibinfo{isbn}{978-0-429-96776-4, 978-0-367-09172-9, 978-0-429-49921-0,
  978-0-7382-0233-4}, \bibinfo{doi}{\doi{10.1201/9780429499210}}.

\bibtype{Book}%
\bibitem{Quigg:2013ufa}
\bibinfo{author}{Chris Quigg}, \bibinfo{title}{{Gauge Theories of the Strong,
  Weak, and Electromagnetic Interactions}: {Second Edition}},
  \bibinfo{publisher}{Princeton University Press}, \bibinfo{address}{USA}
  \bibinfo{year}{2013}, ISBN \bibinfo{isbn}{978-0-691-13548-9,
  978-1-4008-4822-5}, \bibinfo{doi}{\doi{10.1515/9781400848225}}.

\bibtype{Book}%
\bibitem{Weinberg:1995mt}
\bibinfo{author}{Steven Weinberg}, \bibinfo{title}{{The Quantum theory of
  fields. Vol. 1: Foundations}}, \bibinfo{publisher}{Cambridge University
  Press} \bibinfo{year}{2005}, ISBN \bibinfo{isbn}{978-0-521-67053-1,
  978-0-511-25204-4}, \bibinfo{doi}{\doi{10.1017/CBO9781139644167}}.

\bibtype{Book}%
\bibitem{Itzykson:1980rh}
\bibinfo{author}{C. Itzykson}, \bibinfo{author}{J.~B. Zuber},
  \bibinfo{title}{{Quantum Field Theory}}, International Series In Pure and
  Applied Physics, \bibinfo{publisher}{McGraw-Hill}, \bibinfo{address}{New
  York} \bibinfo{year}{1980}, ISBN \bibinfo{isbn}{978-0-486-44568-7}.

\bibtype{Book}%
\bibitem{Cheng:1984vwu}
\bibinfo{author}{Ta-Pei [0000-0002-1137-0969] Cheng},
  \bibinfo{author}{Ling-Fong [0000-0002-8035-3329] Li}, \bibinfo{title}{{Gauge
  Theory of Elementary Particle Physics}}, \bibinfo{publisher}{Oxford
  University Press}, \bibinfo{address}{Oxford, UK} \bibinfo{year}{1984}, ISBN
  \bibinfo{isbn}{978-0-19-851961-4, 978-0-19-851961-4}.

\bibtype{Book}%
\bibitem{Peskin:1995ev}
\bibinfo{author}{Michael~E. Peskin}, \bibinfo{author}{Daniel~V. Schroeder},
  \bibinfo{title}{{An Introduction to quantum field theory}},
  \bibinfo{publisher}{Addison-Wesley}, \bibinfo{address}{Reading, USA}
  \bibinfo{year}{1995}, ISBN \bibinfo{isbn}{978-0-201-50397-5,
  978-0-429-50355-9, 978-0-429-49417-8},
  \bibinfo{doi}{\doi{10.1201/9780429503559}}.

\bibtype{Article}%
\bibitem{Meissner:2024ona}
\bibinfo{author}{Ulf-G. Mei\ss{}ner}, \bibinfo{title}{{Chiral perturbation
  theory}}  (\bibinfo{year}{2024}), \eprint{2410.21912}.

\bibtype{Article}%
\bibitem{Adler:1969gk}
\bibinfo{author}{Stephen~L. Adler}, \bibinfo{title}{{Axial vector vertex in
  spinor electrodynamics}}, \bibinfo{journal}{Phys. Rev.} \bibinfo{volume}{177}
  (\bibinfo{year}{1969}) \bibinfo{pages}{2426--2438},
  \bibinfo{doi}{\doi{10.1103/PhysRev.177.2426}}.

\bibtype{Article}%
\bibitem{Bell:1969ts}
\bibinfo{author}{J.~S. Bell}, \bibinfo{author}{R. Jackiw}, \bibinfo{title}{{A
  PCAC puzzle: $\pi^0 \to \gamma \gamma$ in the $\sigma$ model}},
  \bibinfo{journal}{Nuovo Cim. A} \bibinfo{volume}{60} (\bibinfo{year}{1969})
  \bibinfo{pages}{47--61}, \bibinfo{doi}{\doi{10.1007/BF02823296}}.

\bibtype{Article}%
\bibitem{Bryman:2025pet}
\bibinfo{author}{Douglas Bryman}, \bibinfo{author}{Robert Shrock},
  \bibinfo{title}{{Pion Decay}}  (\bibinfo{year}{2025}), \eprint{2502.18384}.

\bibtype{Book}%
\bibitem{Bertlmann:1996xk}
\bibinfo{author}{R.~A. Bertlmann}, \bibinfo{title}{{Anomalies in quantum field
  theory}} \bibinfo{year}{1996}.

\bibtype{Article}%
\bibitem{tHooft:1973alw}
\bibinfo{author}{Gerard 't Hooft}, \bibinfo{title}{{A Planar Diagram Theory for
  Strong Interactions}}, \bibinfo{journal}{Nucl. Phys. B} \bibinfo{volume}{72}
  (\bibinfo{year}{1974}) \bibinfo{pages}{461},
  \bibinfo{doi}{\doi{10.1016/0550-3213(74)90154-0}}.

\bibtype{Article}%
\bibitem{ParticleDataGroup:2024cfk}
\bibinfo{author}{S. Navas}, et al. (\bibinfo{collaboration}{Particle Data
  Group}), \bibinfo{title}{{Review of particle physics}},
  \bibinfo{journal}{Phys. Rev. D} \bibinfo{volume}{110} (\bibinfo{number}{3})
  (\bibinfo{year}{2024}) \bibinfo{pages}{030001},
  \bibinfo{doi}{\doi{10.1103/PhysRevD.110.030001}}.

\bibtype{Article}%
\bibitem{Gell-Mann:1961omu}
\bibinfo{author}{Murray Gell-Mann}, \bibinfo{title}{{The Eightfold Way: A
  Theory of strong interaction symmetry}}  (\bibinfo{year}{1961}),
  \bibinfo{doi}{\doi{10.2172/4008239}}.

\bibtype{Article}%
\bibitem{Neeman:1961jhl}
\bibinfo{author}{Yuval Ne'eman}, \bibinfo{title}{{Derivation of strong
  interactions from a gauge invariance}}, \bibinfo{journal}{Nucl. Phys.}
  \bibinfo{volume}{26} (\bibinfo{year}{1961}) \bibinfo{pages}{222--229},
  \bibinfo{doi}{\doi{10.1016/0029-5582(61)90134-1}}.

\bibtype{Article}%
\bibitem{Okubo:1961jc}
\bibinfo{author}{Susumu Okubo}, \bibinfo{title}{{Note on unitary symmetry in
  strong interactions}}, \bibinfo{journal}{Prog. Theor. Phys.}
  \bibinfo{volume}{27} (\bibinfo{year}{1962}) \bibinfo{pages}{949--966},
  \bibinfo{doi}{\doi{10.1143/PTP.27.949}}.

\bibtype{Article}%
\bibitem{Goldstone:1961eq}
\bibinfo{author}{J. Goldstone}, \bibinfo{title}{{Field Theories with
  Superconductor Solutions}}, \bibinfo{journal}{Nuovo Cim.}
  \bibinfo{volume}{19} (\bibinfo{year}{1961}) \bibinfo{pages}{154--164},
  \bibinfo{doi}{\doi{10.1007/BF02812722}}.

\bibtype{Article}%
\bibitem{Bernstein:1974rd}
\bibinfo{author}{J. Bernstein}, \bibinfo{title}{{Spontaneous symmetry breaking,
  gauge theories, the higgs mechanism and all that}}, \bibinfo{journal}{Rev.
  Mod. Phys.} \bibinfo{volume}{46} (\bibinfo{year}{1974})
  \bibinfo{pages}{7--48}, \bibinfo{doi}{\doi{10.1103/RevModPhys.46.7}},
  \bibinfo{note}{[Erratum: Rev.Mod.Phys. 47, 259--259 (1975), Erratum:
  Rev.Mod.Phys. 46, 855--855 (1974)]}.

\bibtype{Article}%
\bibitem{Weinberg:1975ui}
\bibinfo{author}{Steven Weinberg}, \bibinfo{title}{{The U(1) Problem}},
  \bibinfo{journal}{Phys. Rev. D} \bibinfo{volume}{11} (\bibinfo{year}{1975})
  \bibinfo{pages}{3583--3593}, \bibinfo{doi}{\doi{10.1103/PhysRevD.11.3583}}.

\bibtype{Article}%
\bibitem{Witten:1980sp}
\bibinfo{author}{Edward Witten}, \bibinfo{title}{{Large N Chiral Dynamics}},
  \bibinfo{journal}{Annals Phys.} \bibinfo{volume}{128} (\bibinfo{year}{1980})
  \bibinfo{pages}{363}, \bibinfo{doi}{\doi{10.1016/0003-4916(80)90325-5}}.

\bibtype{Article}%
\bibitem{DiVecchia:1980yfw}
\bibinfo{author}{P. Di~Vecchia}, \bibinfo{author}{G. Veneziano},
  \bibinfo{title}{{Chiral Dynamics in the Large n Limit}},
  \bibinfo{journal}{Nucl. Phys. B} \bibinfo{volume}{171} (\bibinfo{year}{1980})
  \bibinfo{pages}{253--272}, \bibinfo{doi}{\doi{10.1016/0550-3213(80)90370-3}}.

\bibtype{Article}%
\bibitem{Adler:1964um}
\bibinfo{author}{Stephen~L. Adler}, \bibinfo{title}{{Consistency conditions on
  the strong interactions implied by a partially conserved axial vector
  current}}, \bibinfo{journal}{Phys. Rev.} \bibinfo{volume}{137}
  (\bibinfo{year}{1965}) \bibinfo{pages}{B1022--B1033},
  \bibinfo{doi}{\doi{10.1103/PhysRev.137.B1022}}.

\bibtype{Article}%
\bibitem{Weinberg:1966kf}
\bibinfo{author}{Steven Weinberg}, \bibinfo{title}{{Pion scattering lengths}},
  \bibinfo{journal}{Phys. Rev. Lett.} \bibinfo{volume}{17}
  (\bibinfo{year}{1966}) \bibinfo{pages}{616--621},
  \bibinfo{doi}{\doi{10.1103/PhysRevLett.17.616}}.

\bibtype{Article}%
\bibitem{Nambu:1961tp}
\bibinfo{author}{Yoichiro Nambu}, \bibinfo{author}{G. Jona-Lasinio},
  \bibinfo{title}{{Dynamical Model of Elementary Particles Based on an Analogy
  with Superconductivity. 1.}}, \bibinfo{journal}{Phys. Rev.}
  \bibinfo{volume}{122} (\bibinfo{year}{1961}) \bibinfo{pages}{345--358},
  \bibinfo{doi}{\doi{10.1103/PhysRev.122.345}}.

\bibtype{Article}%
\bibitem{Nambu:1961fr}
\bibinfo{author}{Yoichiro Nambu}, \bibinfo{author}{G. Jona-Lasinio},
  \bibinfo{title}{{Dynamical model of elementary particles based on an analogy
  with superconductivity. II.}}, \bibinfo{journal}{Phys. Rev.}
  \bibinfo{volume}{124} (\bibinfo{year}{1961}) \bibinfo{pages}{246--254},
  \bibinfo{doi}{\doi{10.1103/PhysRev.124.246}}.

\bibtype{Article}%
\bibitem{Bernard:1987sg}
\bibinfo{author}{Veronique Bernard}, \bibinfo{author}{R.~L. Jaffe},
  \bibinfo{author}{Ulf~G. Meissner}, \bibinfo{title}{{Strangeness Mixing and
  Quenching in the Nambu-Jona-Lasinio Model}}, \bibinfo{journal}{Nucl. Phys. B}
  \bibinfo{volume}{308} (\bibinfo{year}{1988}) \bibinfo{pages}{753--790},
  \bibinfo{doi}{\doi{10.1016/0550-3213(88)90127-7}}.

\bibtype{Article}%
\bibitem{Vogl:1991qt}
\bibinfo{author}{U. Vogl}, \bibinfo{author}{W. Weise}, \bibinfo{title}{{The
  Nambu and Jona Lasinio model: Its implications for hadrons and nuclei}},
  \bibinfo{journal}{Prog. Part. Nucl. Phys.} \bibinfo{volume}{27}
  (\bibinfo{year}{1991}) \bibinfo{pages}{195--272},
  \bibinfo{doi}{\doi{10.1016/0146-6410(91)90005-9}}.

\bibtype{Article}%
\bibitem{Klevansky:1992qe}
\bibinfo{author}{S.~P. Klevansky}, \bibinfo{title}{{The Nambu-Jona-Lasinio
  model of quantum chromodynamics}}, \bibinfo{journal}{Rev. Mod. Phys.}
  \bibinfo{volume}{64} (\bibinfo{year}{1992}) \bibinfo{pages}{649--708},
  \bibinfo{doi}{\doi{10.1103/RevModPhys.64.649}}.

\bibtype{Article}%
\bibitem{Hatsuda:1994pi}
\bibinfo{author}{Tetsuo Hatsuda}, \bibinfo{author}{Teiji Kunihiro},
  \bibinfo{title}{{QCD phenomenology based on a chiral effective Lagrangian}},
  \bibinfo{journal}{Phys. Rept.} \bibinfo{volume}{247} (\bibinfo{year}{1994})
  \bibinfo{pages}{221--367}, \bibinfo{doi}{\doi{10.1016/0370-1573(94)90022-1}},
  \eprint{hep-ph/9401310}.

\bibtype{Article}%
\bibitem{Buballa:2003qv}
\bibinfo{author}{Michael Buballa}, \bibinfo{title}{{NJL model analysis of quark
  matter at large density}}, \bibinfo{journal}{Phys. Rept.}
  \bibinfo{volume}{407} (\bibinfo{year}{2005}) \bibinfo{pages}{205--376},
  \bibinfo{doi}{\doi{10.1016/j.physrep.2004.11.004}}, \eprint{hep-ph/0402234}.

\bibtype{Article}%
\bibitem{Volkov:2005kw}
\bibinfo{author}{M.~K. Volkov}, \bibinfo{author}{A.~E. Radzhabov},
  \bibinfo{title}{{The Nambu-Jona-Lasinio model and its development}},
  \bibinfo{journal}{Phys. Usp.} \bibinfo{volume}{49} (\bibinfo{year}{2006})
  \bibinfo{pages}{551--561},
  \bibinfo{doi}{\doi{10.1070/PU2006v049n06ABEH005905}},
  \eprint{hep-ph/0508263}.

\bibtype{Article}%
\bibitem{Amer:1983qa}
\bibinfo{author}{A. Amer}, \bibinfo{author}{A. Le~Yaouanc}, \bibinfo{author}{L.
  Oliver}, \bibinfo{author}{O. Pene}, \bibinfo{author}{J.~c. Raynal},
  \bibinfo{title}{{INSTABILITY OF THE CHIRAL INVARIANT VACUUM FOR A CONFINING
  POTENTIAL}}, \bibinfo{journal}{Phys. Rev. Lett.} \bibinfo{volume}{50}
  (\bibinfo{year}{1983}) \bibinfo{pages}{87--90},
  \bibinfo{doi}{\doi{10.1103/PhysRevLett.50.87}}.

\bibtype{Article}%
\bibitem{LeYaouanc:1983it}
\bibinfo{author}{A. Le~Yaouanc}, \bibinfo{author}{L. Oliver},
  \bibinfo{author}{O. Pene}, \bibinfo{author}{J.~C. Raynal},
  \bibinfo{title}{{Chiral Noninvariant Solutions of the Gap Equation for a
  Confining Potential}}, \bibinfo{journal}{Phys. Lett. B} \bibinfo{volume}{134}
  (\bibinfo{year}{1984}) \bibinfo{pages}{249},
  \bibinfo{doi}{\doi{10.1016/0370-2693(84)90681-6}}.

\bibtype{Article}%
\bibitem{LeYaouanc:1983huv}
\bibinfo{author}{A. Le~Yaouanc}, \bibinfo{author}{L. Oliver},
  \bibinfo{author}{O. Pene}, \bibinfo{author}{J.~C. Raynal},
  \bibinfo{title}{{Spontaneous Breaking of Chiral Symmetry for Confining
  Potentials}}, \bibinfo{journal}{Phys. Rev. D} \bibinfo{volume}{29}
  (\bibinfo{year}{1984}) \bibinfo{pages}{1233--1257},
  \bibinfo{doi}{\doi{10.1103/PhysRevD.29.1233}}.

\bibtype{Article}%
\bibitem{LeYaouanc:1984ntu}
\bibinfo{author}{A. Le~Yaouanc}, \bibinfo{author}{L. Oliver},
  \bibinfo{author}{S. Ono}, \bibinfo{author}{O. Pene}, \bibinfo{author}{J.~C.
  Raynal}, \bibinfo{title}{{A Quark Model of Light Mesons with Dynamically
  Broken Chiral Symmetry}}, \bibinfo{journal}{Phys. Rev. D}
  \bibinfo{volume}{31} (\bibinfo{year}{1985}) \bibinfo{pages}{137--159},
  \bibinfo{doi}{\doi{10.1103/PhysRevD.31.137}}.

\bibtype{Article}%
\bibitem{Adler:1984ri}
\bibinfo{author}{Stephen~L. Adler}, \bibinfo{author}{A.~C. Davis},
  \bibinfo{title}{{Chiral Symmetry Breaking in Coulomb Gauge QCD}},
  \bibinfo{journal}{Nucl. Phys. B} \bibinfo{volume}{244} (\bibinfo{year}{1984})
  \bibinfo{pages}{469}, \bibinfo{doi}{\doi{10.1016/0550-3213(84)90324-9}}.

\bibtype{Article}%
\bibitem{Kocic:1985uq}
\bibinfo{author}{Aleksandar Kocic}, \bibinfo{title}{{Chiral Symmetry
  Restoration at Finite Densities in Coulomb Gauge {QCD}}},
  \bibinfo{journal}{Phys. Rev. D} \bibinfo{volume}{33} (\bibinfo{year}{1986})
  \bibinfo{pages}{1785}, \bibinfo{doi}{\doi{10.1103/PhysRevD.33.1785}}.

\bibtype{Article}%
\bibitem{Bicudo:1989sh}
\bibinfo{author}{Pedro J. de~A. Bicudo}, \bibinfo{author}{Jose E. F.~T.
  Ribeiro}, \bibinfo{title}{{Current Quark Model in a $p$ Wave Triplet
  Condensed Vacuum. 1. The Dynamical Breaking of Chiral Symmetry}},
  \bibinfo{journal}{Phys. Rev. D} \bibinfo{volume}{42} (\bibinfo{year}{1990})
  \bibinfo{pages}{1611--1624}, \bibinfo{doi}{\doi{10.1103/PhysRevD.42.1611}}.

\bibtype{Article}%
\bibitem{Bicudo:1989si}
\bibinfo{author}{Pedro J. de~A. Bicudo}, \bibinfo{author}{Jose E. F.~T.
  Ribeiro}, \bibinfo{title}{{Current Quark Model in a $p$ Wave Triplet
  Condensed Vacuum. 2. Salpeter Equations: $\pi$, $K$, $\rho$, $\phi$ as $q
  \bar{q}$ Bound States}}, \bibinfo{journal}{Phys. Rev. D} \bibinfo{volume}{42}
  (\bibinfo{year}{1990}) \bibinfo{pages}{1625--1634},
  \bibinfo{doi}{\doi{10.1103/PhysRevD.42.1625}}.

\bibtype{Article}%
\bibitem{Bicudo:2002eu}
\bibinfo{author}{P.~J.~A. Bicudo}, \bibinfo{author}{J.~E. F.~T. Ribeiro},
  \bibinfo{author}{A.~V. Nefediev}, \bibinfo{title}{{Vacuum replicas in QCD}},
  \bibinfo{journal}{Phys. Rev. D} \bibinfo{volume}{65} (\bibinfo{year}{2002})
  \bibinfo{pages}{085026}, \bibinfo{doi}{\doi{10.1103/PhysRevD.65.085026}},
  \eprint{hep-ph/0201173}.

\bibtype{Article}%
\bibitem{Llanes-Estrada:1999nat}
\bibinfo{author}{Felipe~J. Llanes-Estrada}, \bibinfo{author}{Stephen~R.
  Cotanch}, \bibinfo{title}{{Meson structure in a relativistic many body
  approach}}, \bibinfo{journal}{Phys. Rev. Lett.} \bibinfo{volume}{84}
  (\bibinfo{year}{2000}) \bibinfo{pages}{1102--1105},
  \bibinfo{doi}{\doi{10.1103/PhysRevLett.84.1102}}, \eprint{hep-ph/9906359}.

\bibtype{Article}%
\bibitem{Bicudo:2003cy}
\bibinfo{author}{P.~J.~A. Bicudo}, \bibinfo{author}{A.~V. Nefediev},
  \bibinfo{title}{{Chiral symmetry breaking solutions for QCD in the truncated
  Coulomb gauge}}, \bibinfo{journal}{Phys. Rev. D} \bibinfo{volume}{68}
  (\bibinfo{year}{2003}) \bibinfo{pages}{065021},
  \bibinfo{doi}{\doi{10.1103/PhysRevD.68.065021}}, \eprint{hep-ph/0307302}.

\bibtype{Article}%
\bibitem{Nefediev:2004by}
\bibinfo{author}{A.~V. Nefediev}, \bibinfo{author}{J.~E. F.~T. Ribeiro},
  \bibinfo{title}{{Mesonic states and vacuum replicas in potential quark models
  for QCD}}, \bibinfo{journal}{Phys. Rev. D} \bibinfo{volume}{70}
  (\bibinfo{year}{2004}) \bibinfo{pages}{094020},
  \bibinfo{doi}{\doi{10.1103/PhysRevD.70.094020}}, \eprint{hep-ph/0409112}.

\bibtype{Article}%
\bibitem{Alkofer:2005ug}
\bibinfo{author}{Reinhard Alkofer}, \bibinfo{author}{M. Kloker},
  \bibinfo{author}{A. Krassnigg}, \bibinfo{author}{R.~F. Wagenbrunn},
  \bibinfo{title}{{Aspects of the confinement mechanism in Coulomb-gauge QCD}},
  \bibinfo{journal}{Phys. Rev. Lett.} \bibinfo{volume}{96}
  (\bibinfo{year}{2006}) \bibinfo{pages}{022001},
  \bibinfo{doi}{\doi{10.1103/PhysRevLett.96.022001}}, \eprint{hep-ph/0510028}.

\bibtype{Article}%
\bibitem{Wagenbrunn:2007ie}
\bibinfo{author}{R.~F. Wagenbrunn}, \bibinfo{author}{L.~Ya. Glozman},
  \bibinfo{title}{{Chiral symmetry patterns of excited mesons with the
  Coulomb-like linear confinement}}, \bibinfo{journal}{Phys. Rev. D}
  \bibinfo{volume}{75} (\bibinfo{year}{2007}) \bibinfo{pages}{036007},
  \bibinfo{doi}{\doi{10.1103/PhysRevD.75.036007}}, \eprint{hep-ph/0701039}.

\bibtype{Article}%
\bibitem{Bicudo:2008kc}
\bibinfo{author}{P. Bicudo}, \bibinfo{title}{{Chiral symmetry breaking in the
  truncated Coulomb Gauge. II. Non-confining power law potentials}},
  \bibinfo{journal}{Phys. Rev. D} \bibinfo{volume}{79} (\bibinfo{year}{2009})
  \bibinfo{pages}{094030}, \bibinfo{doi}{\doi{10.1103/PhysRevD.79.094030}},
  \eprint{0811.0407}.

\bibtype{Article}%
\bibitem{tHooft:1974pnl}
\bibinfo{author}{Gerard 't Hooft}, \bibinfo{title}{{A Two-Dimensional Model for
  Mesons}}, \bibinfo{journal}{Nucl. Phys. B} \bibinfo{volume}{75}
  (\bibinfo{year}{1974}) \bibinfo{pages}{461--470},
  \bibinfo{doi}{\doi{10.1016/0550-3213(74)90088-1}}.

\bibtype{Article}%
\bibitem{Bars:1977ud}
\bibinfo{author}{I. Bars}, \bibinfo{author}{Michael~B. Green},
  \bibinfo{title}{{Poincare and Gauge Invariant Two-Dimensional QCD}},
  \bibinfo{journal}{Phys. Rev. D} \bibinfo{volume}{17} (\bibinfo{year}{1978})
  \bibinfo{pages}{537}, \bibinfo{doi}{\doi{10.1103/PhysRevD.17.537}}.

\bibtype{Article}%
\bibitem{Li:1986gf}
\bibinfo{author}{M. Li}, \bibinfo{title}{{Large $N$ Two-dimensional {QCD} and
  Chiral Symmetry}}, \bibinfo{journal}{Phys. Rev. D} \bibinfo{volume}{34}
  (\bibinfo{year}{1986}) \bibinfo{pages}{3888--3893},
  \bibinfo{doi}{\doi{10.1103/PhysRevD.34.3888}}.

\bibtype{Article}%
\bibitem{Kalashnikova:2001df}
\bibinfo{author}{Yu.~S. Kalashnikova}, \bibinfo{author}{A.~V. Nefediev},
  \bibinfo{title}{{Two-dimensional QCD in the Coulomb gauge}},
  \bibinfo{journal}{Phys. Usp.} \bibinfo{volume}{45} (\bibinfo{year}{2002})
  \bibinfo{pages}{347--368},
  \bibinfo{doi}{\doi{10.1070/PU2002v045n04ABEH001070}},
  \eprint{hep-ph/0111225}.

\bibtype{Article}%
\bibitem{Glozman:2012ev}
\bibinfo{author}{L.~Ya. Glozman}, \bibinfo{author}{V.~K. Sazonov},
  \bibinfo{author}{M. Shifman}, \bibinfo{author}{R.~F. Wagenbrunn},
  \bibinfo{title}{{How Chiral Symmetry Breaking Affects the Spectrum of the
  Light-Heavy Mesons in the 't Hooft Model}}, \bibinfo{journal}{Phys. Rev. D}
  \bibinfo{volume}{85} (\bibinfo{year}{2012}) \bibinfo{pages}{094030},
  \bibinfo{doi}{\doi{10.1103/PhysRevD.85.094030}}, \eprint{1201.5814}.

\bibtype{Article}%
\bibitem{Christ:1980ku}
\bibinfo{author}{N.~H. Christ}, \bibinfo{author}{T.~D. Lee},
  \bibinfo{title}{{Operator Ordering and Feynman Rules in Gauge Theories}},
  \bibinfo{journal}{Phys. Rev. D} \bibinfo{volume}{22} (\bibinfo{year}{1980})
  \bibinfo{pages}{939}, \bibinfo{doi}{\doi{10.1103/PhysRevD.22.939}}.

\bibtype{Article}%
\bibitem{Szczepaniak:2001rg}
\bibinfo{author}{Adam~P. Szczepaniak}, \bibinfo{author}{Eric~S. Swanson},
  \bibinfo{title}{{Coulomb gauge QCD, confinement, and the constituent
  representation}}, \bibinfo{journal}{Phys. Rev. D} \bibinfo{volume}{65}
  (\bibinfo{year}{2001}) \bibinfo{pages}{025012},
  \bibinfo{doi}{\doi{10.1103/PhysRevD.65.025012}}, \eprint{hep-ph/0107078}.

\bibtype{Article}%
\bibitem{Feuchter:2004mk}
\bibinfo{author}{C. Feuchter}, \bibinfo{author}{H. Reinhardt},
  \bibinfo{title}{{Variational solution of the Yang-Mills Schrodinger equation
  in Coulomb gauge}}, \bibinfo{journal}{Phys. Rev. D} \bibinfo{volume}{70}
  (\bibinfo{year}{2004}) \bibinfo{pages}{105021},
  \bibinfo{doi}{\doi{10.1103/PhysRevD.70.105021}}, \eprint{hep-th/0408236}.

\bibtype{Article}%
\bibitem{Reinhardt:2017pyr}
\bibinfo{author}{H. Reinhardt}, \bibinfo{author}{G. Burgio},
  \bibinfo{author}{D. Campagnari}, \bibinfo{author}{E. Ebadati},
  \bibinfo{author}{J. Heffner}, \bibinfo{author}{M. Quandt},
  \bibinfo{author}{P. Vastag}, \bibinfo{author}{H. Vogt},
  \bibinfo{title}{{Hamiltonian approach to QCD in Coulomb gauge - a survey of
  recent results}}, \bibinfo{journal}{Adv. High Energy Phys.}
  \bibinfo{volume}{2018} (\bibinfo{year}{2018}) \bibinfo{pages}{2312498},
  \bibinfo{doi}{\doi{10.1155/2018/2312498}}, \eprint{1706.02702}.

\bibtype{Article}%
\bibitem{Nguyen:2024ikq}
\bibinfo{author}{Mendel Nguyen}, \bibinfo{author}{Tin Sulejmanpasic},
  \bibinfo{author}{Mithat \"Unsal}, \bibinfo{title}{{Phases of Theories with ZN
  1-Form Symmetry, and the Roles of Center Vortices and Magnetic Monopoles}},
  \bibinfo{journal}{Phys. Rev. Lett.} \bibinfo{volume}{134}
  (\bibinfo{number}{14}) (\bibinfo{year}{2025}) \bibinfo{pages}{141902},
  \bibinfo{doi}{\doi{10.1103/PhysRevLett.134.141902}}, \eprint{2401.04800}.

\bibtype{Article}%
\bibitem{Kalashnikova:2017ssy}
\bibinfo{author}{Yu.~S. Kalashnikova}, \bibinfo{author}{A.~V. Nefediev},
  \bibinfo{author}{J.~E. F.~T. Ribeiro}, \bibinfo{title}{{Chiral symmetry and
  the properties of hadrons in the generalized Nambu - Jona-Lasinio model}},
  \bibinfo{journal}{Phys. Usp.} \bibinfo{volume}{60} (\bibinfo{number}{7})
  (\bibinfo{year}{2017}) \bibinfo{pages}{667--693},
  \bibinfo{doi}{\doi{10.3367/UFNe.2016.11.037966}}, \eprint{1707.04886}.

\bibtype{Article}%
\bibitem{Llewellyn-Smith:1969bcu}
\bibinfo{author}{C.~H. Llewellyn-Smith}, \bibinfo{title}{{A relativistic
  formulation for the quark model for mesons}}, \bibinfo{journal}{Annals Phys.}
  \bibinfo{volume}{53} (\bibinfo{year}{1969}) \bibinfo{pages}{521--558},
  \bibinfo{doi}{\doi{10.1016/0003-4916(69)90035-9}}.

\bibtype{Article}%
\bibitem{GellMann:1968rz}
\bibinfo{author}{Murray Gell-Mann}, \bibinfo{author}{R.~J. Oakes},
  \bibinfo{author}{B. Renner}, \bibinfo{title}{{Behavior of current divergences
  under SU(3) x SU(3)}}, \bibinfo{journal}{Phys. Rev.} \bibinfo{volume}{175}
  (\bibinfo{year}{1968}) \bibinfo{pages}{2195--2199},
  \bibinfo{doi}{\doi{10.1103/PhysRev.175.2195}}.

\bibtype{Article}%
\bibitem{Eichmann:2025wgs}
\bibinfo{author}{Gernot Eichmann}, \bibinfo{title}{{Hadron physics with
  functional methods}}  (\bibinfo{year}{2025}), \eprint{2503.10397}.

\bibtype{Book}%
\bibitem{Meissner:2022cbi}
\bibinfo{author}{Ulf-G. Mei\ss{}ner}, \bibinfo{author}{Akaki Rusetsky},
  \bibinfo{title}{{Effective Field Theories}}, \bibinfo{publisher}{Cambridge
  University Press} \bibinfo{year}{2022}, ISBN
  \bibinfo{isbn}{978-1-108-68903-8},
  \bibinfo{doi}{\doi{10.1017/9781108689038}}.

\end{thebibliography*}

%
%

\end{document}